\documentclass[astrosymb, twocolumn, tighten, twocolappendix]{aastex631}
\usepackage{amsmath,amssymb}
\usepackage[utf8]{inputenc}
\usepackage{graphicx}
\usepackage{mathptmx,txfonts,tikz,bm,booktabs,multirow}
\usepackage[capitalize]{cleveref}

\graphicspath{{figures/},{./}}

\usepackage{CJKutf8}
\newcommand{\chinesename}{{\begin{CJK}{UTF8}{gbsn}(王加冕)\end{CJK}}}

\begin{document}

\title{Evolutionary and Observational Properties of Red Giant Acoustic Glitch Signatures}
\author[0000-0001-8162-8227]{David P. Saunders}
\affiliation{Department of Astronomy, Yale University, PO Box 208101, New Haven, CT 06520-8101, USA}

\author[0000-0001-7664-648X]{J. M. Joel Ong \chinesename}
\affiliation{Department of Astronomy, Yale University, PO Box 208101, New Haven, CT 06520-8101, USA}
% \affiliation{Hubble Fellow}

\author[0000-0002-6163-3472]{Sarbani Basu}
\affiliation{Department of Astronomy, Yale University, PO Box 208101, New Haven, CT 06520-8101, USA}
\shorttitle{Red Giant Glitches}
\shortauthors{Saunders, Ong, \& Basu}

\newcommand{\numax}{{\ensuremath{\nu_{\text{max}}}}}
\newcommand{\Gone}{{\ensuremath{\Gamma_1}}}
\newcommand{\annotate}[2]{\begin{tikzpicture}
    \node[anchor=south west,inner sep=0,align=center] (image) at (0,0) {
    #1
    };
    \begin{scope}[x={(image.south east)},y={(image.north west)}]
    #2
    \end{scope}
\end{tikzpicture}}

% turn off tracking of changes for n < N0
\renewcommand{\edit}[2]{{\ifnum#1<4%
#2%
\else%
\textbf{#2}%
\fi}}

\begin{abstract}
    While solar-like oscillations in red giants have been observed at massive scale by the \emph{Kepler} mission, few features of these oscillation mode frequencies, other than their global properties, have been exploited for stellar characterization. The signatures of acoustic glitches in mode frequencies have been used for studying main-sequence stars, but the \edit1{validity of applying} such techniques to evolved red giants, particularly pertaining to the inclusion of nonradial modes, has been less well-examined. Making use of new theoretical developments, we characterize glitches using the $\pi$ modes associated with red giant stellar models, and use our procedure to examine for the first time how properties of the He II acoustic glitch --- specifically its amplitude and associated acoustic depth --- vary over the course of evolution up the red giant branch, and with respect to other fundamental stellar properties. We find that the acoustic depths of these glitches, in conjunction with other spectroscopic information, discriminates between red giants in the first-ascent and core-helium-burning phases. We critically reexamine previous attempts to constrain acoustic glitches from nonradial (in particular dipole) modes in red giants. Finally, we apply our fitting procedure to \textit{Kepler} data, to evaluate its robustness to noise and other observational systematics.
\end{abstract}


\section{introduction}

While data from the \emph{Kepler} mission have yielded voluminous asteroseismic observations for red giants, analysis of this asteroseismology has so far been largely limited to catalogs of global seismic and spectroscopic parameters. This in turn has proven its worth, e.g. by permitting differentiation between first-ascent red giant branch (RGB) and red clump (RC) stars, which are otherwise observationally similar \citep[e.g.,][]{2010ApJ...713L.176B,2018ApJS..239...32P,2018ApJS..236...42Y}. However, these seismic observations provide more information than are encapsulated in the global parameters, and this information has yet to be exploited at a similar scale.

Red giant stars behave as solar-like oscillators, exhibiting stochastically-excited modes of oscillation. Some of these modes are acoustic (pressure) modes, which can be described by a comb-like eigenvalue equation:
\begin{equation}
    \nu_{n\ell} \sim \Delta\nu \left(n + {\ell \over 2} + \epsilon_{\ell, p}(\nu)\right). \label{eq:p}
\end{equation}
The quantities $\Delta\nu$ and $\epsilon_p$ (considered as an averaged constant value) are global properties that summarize the overall structure of this comb. While $\epsilon_p$ is indeed close to constant for very simple stellar structures, the actual mode frequencies observed in solar-like oscillators exhibit minute deviations from a strict frequency comb. These deviations between the predicted comb structure and the actual frequencies may, to first order in perturbation theory, be described as combinations of oscillatory components. These components are the signature of `glitches', which are sharp variations in the adiabatic sound speed within the stellar structure. The apparently oscillatory morphology of such glitches lends them easily to being modelled by sinusoidal functions with varying amplitudes \citep[cf. e.g][and references therein]{2014ApJ...794..114V}. Within such descriptive frameworks, glitch signatures may be approximately specified using phenomenological parameters, such as the local amplitude and period of the apparently sinusoidal signature, which may then be used to indirectly constrain the stellar structure. When the glitch lies close to the stellar surface, the period $P$ has been shown to be related to the acoustic depth $\tau$ of the corresponding localized variations in the sound-speed profile \citep[cf.][]{1990LNP...367..283G} as
\begin{equation}
    \tau = \frac{1}{2P} = \int^R_{r_\mathrm{glitch}} {\mathrm d r \over c_s}\label{eq:depth}
\end{equation}
where $R$ is the stellar radius, $c_s$ is the adiabatic sound speed, and $r_\mathrm{glitch}$ is the radial position of the glitch feature. Thus, measurements of the morphology of the observed glitches permit the locations of features in the sound speed profile to be inferred. Since the adiabatic sound speed $c_s$ is tied to the first adiabatic index $\Gamma_1$, knowing the acoustic depth allows for an understanding of the variations in the thermal structure of the star as well. Variations in $c_s$ thus correspond to distinct features of the $\Gamma_1$ profile. There are two kinds of glitches in the sound speed profile which are pertinent to discussions of solar-like oscillators: those arising from boundaries between convective and radiative regions, as well as depressions in $\Gamma_1$ at ionization zones (notably the H~I/He~I and He II ionization zones). 

%The process of ionization leaves a depression in the first adiabatic index $\Gamma_1$, and thus these glitches are visible in the thermal structure of the star. The H~I/He I glitch is too shallow and its signature on the frequencies is basically a smooth function of frequency, and consequently we focus on the He II glitch.  

\edit1{Whereas these characterizations of acoustic glitches were originally developed for describing main-sequence stars}, methodological complications arise when extending these methods to evolved solar-like oscillators. In red giant oscillators, the above description of acoustic modes serves well for both the observed radial and quadrupole modes, but does not for the observed dipole modes; those exhibit mixed character instead. These mixed dipole modes arise when core-bound gravity waves couple to pressure waves in the envelope \citep[e.g.][]{osaki_nonradial_1975,aizenman_avoided_1977}. Pure gravity modes are known to satisfy a separate asymptotic relation,
\begin{equation}
     {1 \over \nu_{n\ell}} \sim \Delta\Pi_\ell \left(n + {\ell \over 2} + \epsilon_{\ell,g}(\nu)\right),\label{eq:g}
\end{equation}
associated with a period spacing $\Delta\Pi_l$ and gravity-mode phase offset $\epsilon_g$, in an analogous fashion to \cref{eq:p}. However, modes of such strongly mixed character as are observed are not well described by either \cref{eq:p} or \cref{eq:g}. As such, the mixed nature of these modes makes them difficult to use even for determining these reduced set of phenomenological glitch parameters: while only the acoustic components of these modes are affected by the glitch, this information is not easily extracted from the observational set of mixed modes.  Even the nominally p-like quadrupole mixed modes of red giant models, which do satisfy \cref{eq:p} well, have hitherto been accessible only at significant, and in many cases prohibitive, computational expense.

\edit1{Consequently, the accuracy of observational prescriptions for constraining pure acoustic glitches from mixed modes has so far not been well-interrogated. The theoretical relationships between these acoustic signatures and the interior structures of mixed-mode oscillators has also not been subjected to nearly the same amount of scrutiny as compared to over the course of their development for main-sequence stars. These considerations are of critical importance in light of ongoing observational efforts to apply glitch characterisation to red giants \cite[e.g.][]{2015A&A...579A..84V,2021A&A...650A.115D}, notwithstanding such unresolved open questions.}

These theoretical and computational difficulties have been alleviated by recent \edit1{analytic} developments. In particular, \cite{ong_semianalytic_2020} \cite[following][]{2018MNRAS.478.4697B} provide a prescription by which the notional pure p-modes of a stellar model \cite[$\pi$-modes, in the sense of ][]{aizenman_avoided_1977} may be recovered, which significantly reduces the computational burden of evaluating the frequencies of the p-dominated quadrupole modes. Access to such pure p-modes also permits us to critically examine previously claimed improvements to the technique, stemming from proposals for deriving pure dipole p-modes from the observed mixed modes \citep[e.g.][]{2020MNRAS.497.1008D}.

% This is impacted by the choice of the stellar acoustic radius T (): 
% \begin{equation}
%     T \approx \frac{1}{2\Delta\nu} \approx \int_0^R \frac{dr}{c_s} % Dreau for 2nd 
% \end{equation}
% where $\Delta\nu$ is the large frequency separation, R is the seismic radius and $c_s$ is the adiabatic sound speed. The calculation using $\Delta\nu$ gives a different result for $\tau$ than that which computes the acoustic radius. The acoustic depth serves to highlight the location of the glitch in the star; however, the imprecise boundary definitions of the acoustic radius make the localization of the glitch difficult. Nevertheless, the calculation of acoustic depth from the large frequency separation $\Delta\nu$ has been shown to give a more accurate value of $\tau$ (). This result is confirmed by the work done in this study. 

In this paper, we examine the potential for using the properties of the helium glitch for constraining some properties of evolved (in particular first-ascent red giant) solar-like oscillators, \edit2{using techniques inherited from the study of these glitches in main-sequence stars}. In \autoref{sec:methods} we discuss the fitting procedure and the chosen parametric model. We use said procedure in \autoref{sec:evol} to examine the relationships between spectroscopic stellar properties and seismic parameters, as well as how the fitted model localizes the He II glitch within the adiabatic structure of the model. In \autoref{sec:disc}, we assess the benefits of including dipole modes in this procedure, and in \autoref{sec:obs} we consider the sensitivity of this procedure to observational uncertainties. Finally, in \autoref{sec:conc} we summarize our key results and potential follow-up work. 

\section{Methods}
\label{sec:methods}

We first implemented an automated procedure with a restricted parameterization adapted for use on red giants, which we benchmark on mode frequencies returned from evolutionary models. The development of this fitting pipeline anchors our current study; all of our results (and subsequent analysis of \emph{Kepler} data) rely upon it. We describe in particular our selection of a specific parameterization to fit the glitch signature, the imposition of various cutoffs for numerical conditioning, and numerical optimization. 

\subsection{Parameterization}

The He II glitch may be characterized via second differences of the frequencies, which are taken in order to reduce the impact of slowly varying components and isolate the oscillatory signal. For an input set of mode frequencies, our pipeline computes these second differences in the usual fashion as in \citet{1990LNP...367..283G}
\begin{equation}
    \delta^2\nu_{n,\ell} = \nu_{n+1,\ell} - 2\nu_{n,\ell} + \nu_{n-1,\ell} \label{eq:seconddiff}. % Verma et al.
\end{equation}
Various parameterizations of these second differences have been proposed in the literature \citep[see, e.g.,][]{2014ApJ...794..114V}. Existing parameterizations generally include two sinusoidal components: an interior term with more rapid oscillations to describe an acoustic glitch at the base of the convection zone, and an exterior term with slower oscillations to describe the He II glitch. However, the morphology of these red giants differs significantly from those of the main-sequence stars for which these parameterizations were developed; in particular, red giants exhibit very compact radiative cores. Consequently, the periodicity of their convective-boundary glitch signatures yields not their acoustic depths, as in \cref{eq:depth}, but rather their acoustic radii (i.e. with the integral limits going from the center of the star to the convective boundary), which are small \citep[see][]{2001A&A...377..192M}. Accordingly, when adapting existing parameterizations to red giants, we omitted terms corresponding to the convective boundary, whose slow variations are effectively detrended away by other terms in these parameterizations.

We tested several parametrizations for the He II term from \citet{basu_asteroseismic_2004} and \cite{2014ApJ...794..114V} and adopt a Gaussian-envelope model, which has the lowest number of parameters, to fit the oscillatory glitch signature in our subsequent analysis: 
\begin{equation}
    G (\nu) = A \nu\,\exp \left[{-b^2 \left(\frac{\nu}{\numax}\right)^2}\right] \cos( 2 \pi \nu/P + \phi) + F(\nu), \label{eq:parameterisation} % Verma et al.  
\end{equation}
where
\begin{equation}
    F(\nu) = c + d\left(\frac{\nu}{\numax}-1\right) + g\left(\frac{\nu}{\numax}-1\right)^2
\end{equation}
is a slowly-varying component. This is included to account for the frequency variations arising from the glitch produced at the base of the convection zone, as well as other smooth components arising from core- and surface-boundary effects also seen in main-sequence stars\edit2{, as well as higher-order terms ordinarily neglected in the asymptotic expansion for the p-mode frequencies}. We found that while the fits were significantly improved by accounting for this slow component, additional terms of degree higher than 2 did not further improve the quality of the fit. 

\subsection{Fitting Process}

Given a set of measurement errors on the mode frequencies, a best-fitting model with respect to our parameterization may be found by minimizing the cost function
\begin{equation}
    \chi^2 = \sum_{i, j}\left(\delta^2\nu_i - G(\nu_i)\right) C^{-1}_{ij}\left(\delta^2\nu_j - G(\nu_j)\right),
\end{equation}
where $C^{-1}$ is the inverse covariance matrix for the second differences, found by propagating uncertainties in the mode frequencies. In principle, uncertainties for the fitting parameters may be determined from the Hessian matrix of this cost function. However, given the highly nonlinear structure of the optimization problem, we instead elected to determine uncertainties in the glitch amplitude by a Monte-Carlo bootstrapping approach, in which the fit to obtain the parameters was performed repeatedly under many realizations of random perturbations to the input data specified by the measurement uncertainties. The final reported uncertainties were taken as the sample standard deviations of the fitted parameters across 100 realizations. We found that using such a Monte-Carlo procedure rendered our results essentially insensitive to the off-diagonal elements of the covariance matrices; hence, we restricted ourselves to only the diagonal elements to accelerate the computation.

When fitting to these second differences, we restricted our mode sets to second differences within a frequency interval $6\Delta\nu$ wide, centered on $\numax$. We implemented this by weighting modes in the fitting procedure with a soft cutoff function, via $\tanh$ functions centered at $ \pm3\Delta\nu$, with a softening length scale of $0.1\Delta\nu$; we did this so as to avoid discontinuous behavior over the course of evolution as modes enter and leave this fixed window, which is the case with a hard cutoff (i.e. giving modes weights of either 0 or 1). No attempt to fit a glitch function was made when the number of second differences within this interval was less than the number of free parameters. 

\begin{figure}[htbp]
    \centering
    \includegraphics[width=8.5cm]{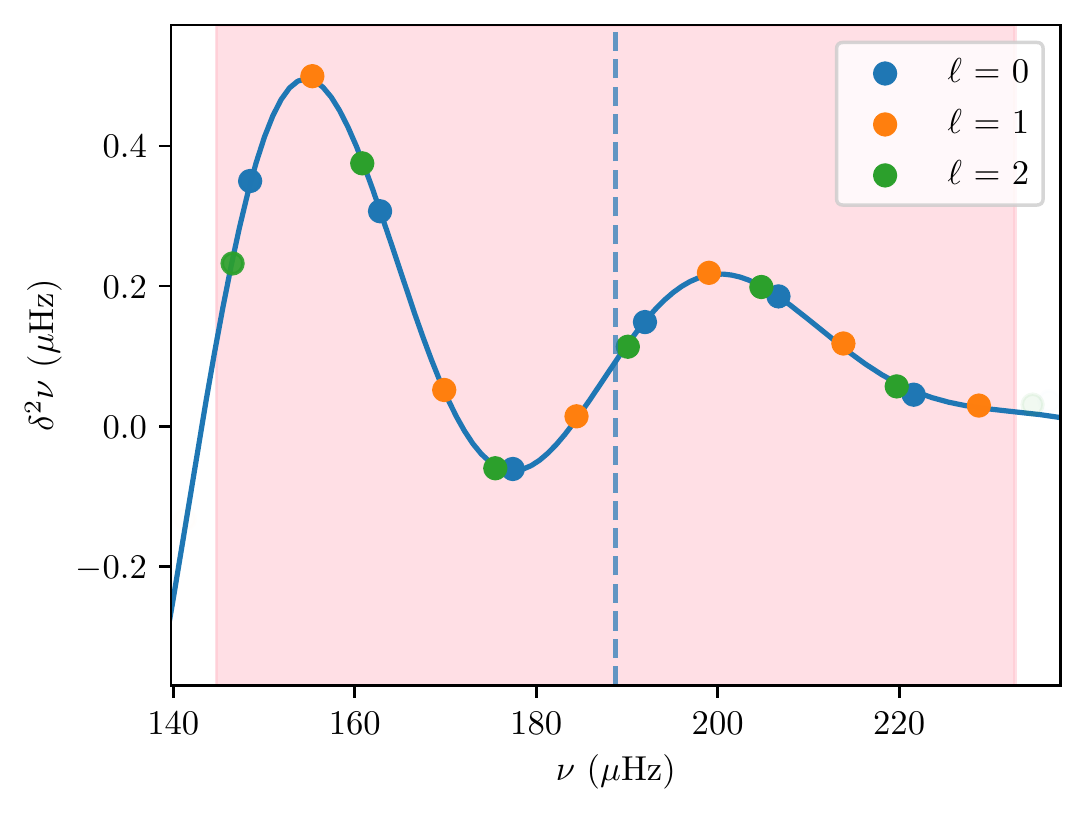}
    \caption{Second differences of a sample glitch signature fitted against frequency. Colored circles indicate the $\ell = 0, 1, 2$ modes of the MESA model star. The solid curve shows the fit to all three of these modes. The dotted line represents the location of $\numax$ on which the fit is centered. The shaded region illustrates the $\pm3\Delta\nu$ area from which modes were utilized in the fit.}
    \label{fig:seconddiffs}
\end{figure}

In certain cases, we found that fits would get stuck in local optima (e.g. at aliases of the true glitch period). We found that, along each evolutionary track, this problem can be somewhat alleviated by using the fitted parameters from the previous timestep as initial guesses in the fitting process. However, doing so would not be an option when confronted with observational data. As a more general means of avoiding local optima, we used the differential evolution algorithm for gradient-free optimization, as implemented in the python package \texttt{yabox} \citep{2017zndo....848679M}. As a further precaution against local optima, we perform an explicit parameter sweep over a grid of possible period values, since the period is the most ill-conditioned parameter. The period grid is bound by the total frequency range on one end, and on the other by a demand that the acoustic depth of the helium glitch be in the outer half of the star:
\begin{equation}
    \tau \leq \frac{1}{4\Delta\nu}.\label{eq:outer}
\end{equation}
The additional parameters of our glitch model, \cref{eq:parameterisation}, are fitted independently with the glitch period held fixed at each point in the grid, and the grid period which produced the lowest $\chi^2$ is chosen to seed a final optimization run where the period, too, is permitted to vary freely. We show a sample result from this procedure in \cref{fig:seconddiffs}. 

\section{Results on Evolutionary Models}
\label{sec:evol}

\edit1{We now seek to understand the relationships between spectroscopic and glitch parameters over the course of stellar evolution, by applying this glitch fitting procedure to synthetic stellar models. For this purpose, we construct a grid of evolutionary models, and use the above procedure to conduct a parameter study of how the helium abundance, metallicity, and stellar mass each affect the amplitudes and periods of the fitted glitch signature.}

\subsection{Parameter Grid}

We generated evolutionary tracks of red giant stellar models using MESA \citep{mesa_paper_1,mesa_paper_2,mesa_paper_3,mesa_paper_4,mesa_paper_5} with element diffusion and a small amount of step convective overshoot ($f_\mathrm{step}$ = 0.0016). Stellar models were first generated on a coarsely equisampled grid of input parameters over the ranges $1 M_\odot \le M \le 2 M_\odot$ (in steps of $0.2 M_\odot$), $0.25 \le Y_0 \le 0.3$ (in steps of 0.025), and $[\mathrm{Fe/H}]_0 \in \{-0.30, 0, +0.30\}$ dex. We consider stellar models from $N_1 = \Delta\nu / \numax^2 \Delta\Pi_1 = 5$ up to core helium exhaustion (at the end of the core helium-burning phase). Mode frequencies were generated using the pulsation code GYRE \citep{townsend_gyre_2013}, \edit1{with the nonradial modes evaluated} according to the $\pi$-mode prescription of \cite{ong_semianalytic_2020}. The glitch fitting algorithm was run \edit1{for each stellar model (i.e. at every timestep)} to relate the glitch amplitude and period to the \edit1{global properties} of the model. 

We supplemented this coarse grid with a further set of evolutionary tracks with much finer sampling, with perturbations to the values of $M_i$, $Y_i$, and [Fe/H]$_i$ intended to match the uncertainties in these parameters typically reported from stellar modelling. The input parameters were sampled at $M = 1.2 M_\odot \pm 2.5\%$, $Y_0 = 0.275 \pm 0.0125$, and $[\mathrm{Fe/H}]_0 = 0 \pm 0.08$ dex. We use this ``fine'' grid to assess \edit1{how} the errors in our glitch-fitting amplitudes, were they to be used as inputs to stellar modelling, would otherwise compare with the variations associated with propagating the uncertainties on spectroscopic stellar parameters.

Frequencies of stellar models do not have any uncertainties. For the sake of further discussion, in order to make a quantitatively commensurate comparison between our results and real data, we assign artificial measurement errors to the mode frequencies using fixed frequency measurement error of $0.01\mu$Hz, representative of a typical frequency measurement error in these quantities, and similarly adopted by \cite{2020MNRAS.497.1008D}.

\subsection{Results}

\edit1{Similar parameter studies conducted for main sequence stars \citep[e.g.][]{2014ApJ...794..114V} have considered the relation between the properties of the glitch, and the global stellar properties $M$, $Y$, and [Fe/H], as well as the internal thermodynamic structures of stellar models. We therefore examine the relationships between the same stellar properties as in those studies, and the glitch amplitudes and periods returned by our glitch fitting pipeline.}

\begin{figure}
    \centering
    \fig{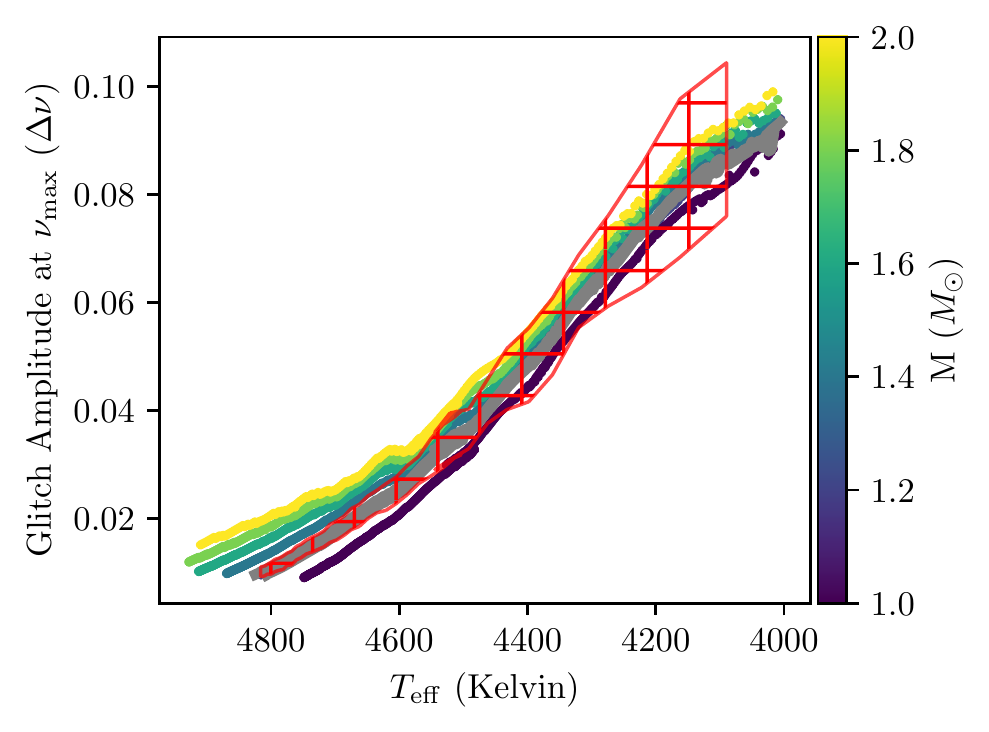}{.45\textwidth}{(a)}
    \fig{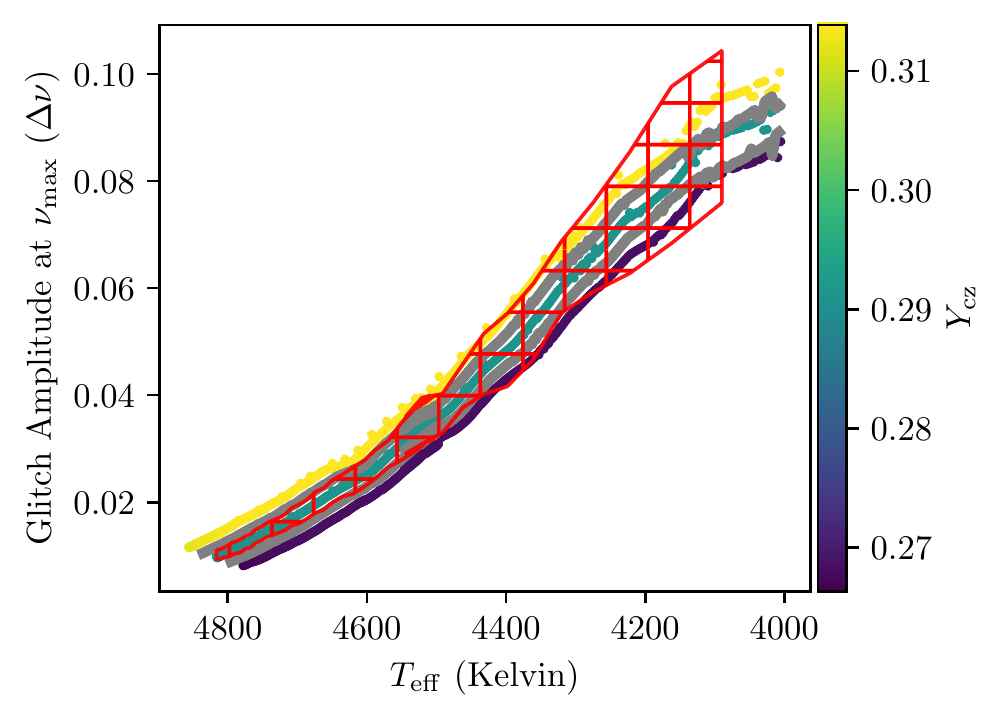}{.45\textwidth}{(b)}
    \fig{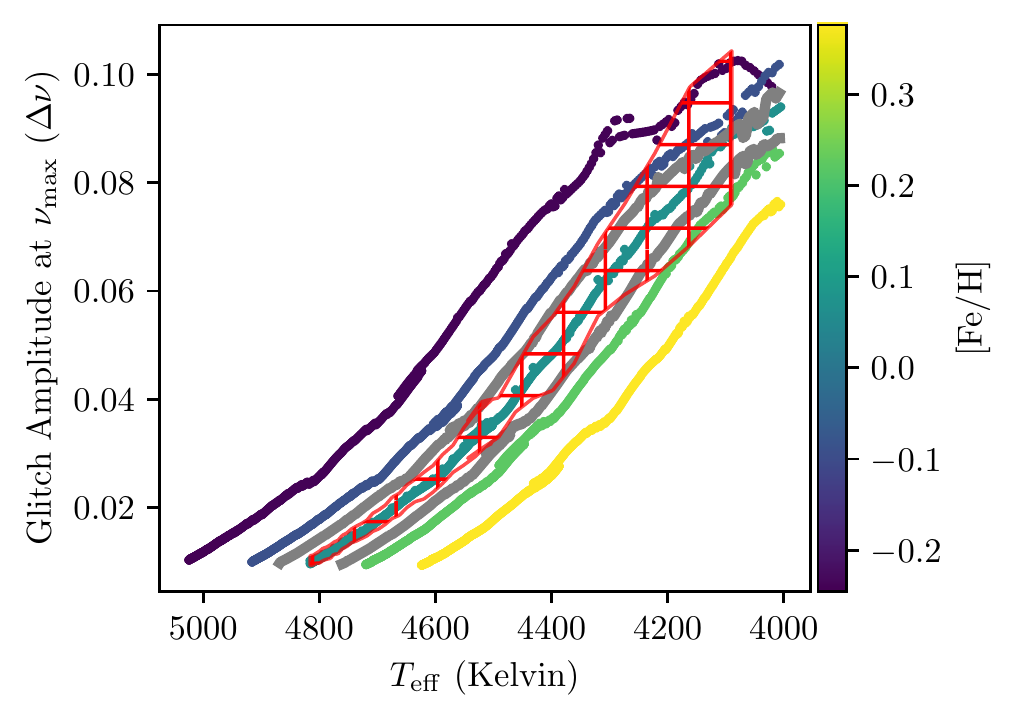}{.45\textwidth}{(c)}
    \caption{Fitted glitch amplitude at $\numax$ plotted against effective temperature, for series of evolutionary tracks of varying \textbf{(a)} initial mass; \textbf{(b)} $Y_\mathrm{CZ}$; \textbf{(c)} [Fe/H]. The red crosshatching represents the glitch amplitude uncertainty calculated on the central track. The grey lines represent the models from the ``fine'' grid, which approximate differences in spectroscopic parameters representative of observational errors.}
    \label{fig:amp_evol}
\end{figure}

\begin{figure}
    \centering
    \fig{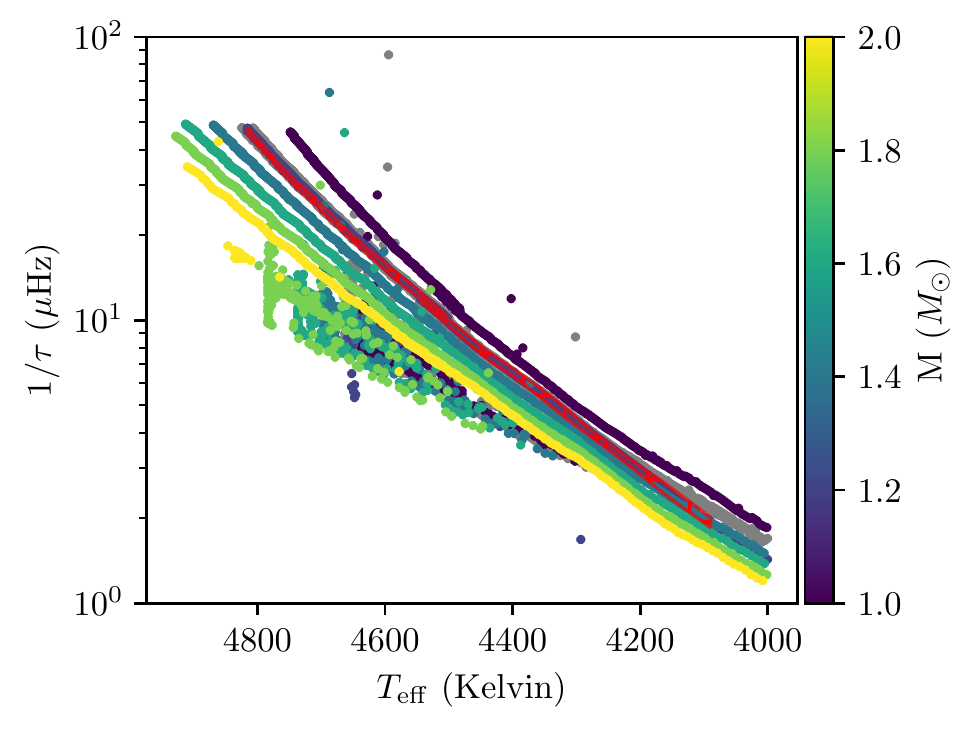}{.45\textwidth}{(a)}
    \fig{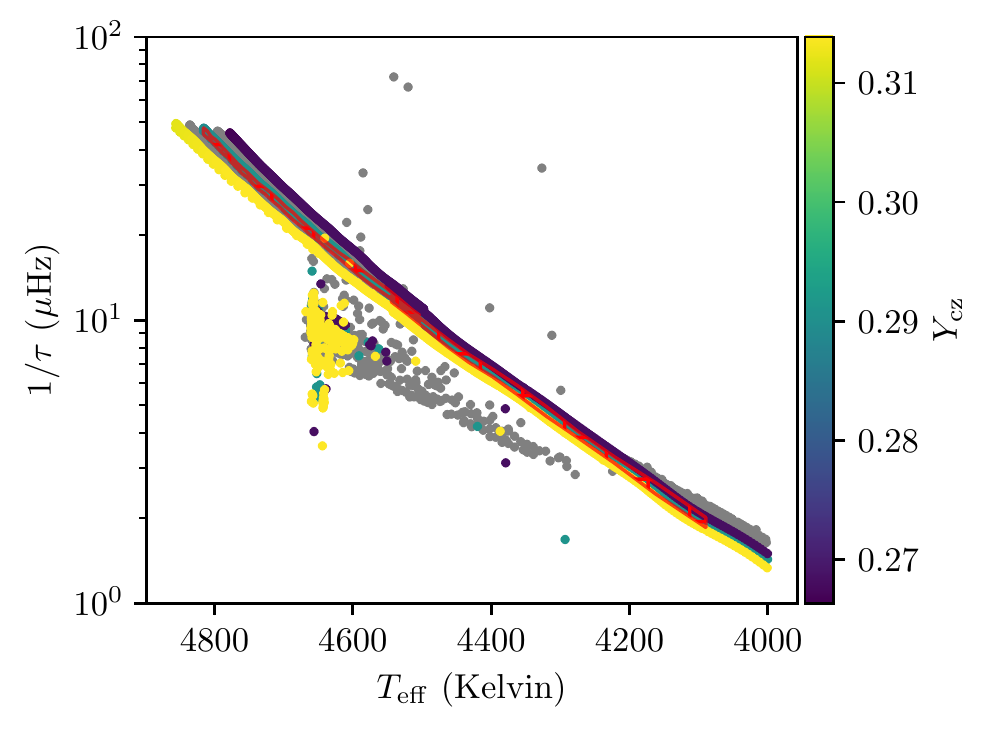}{.45\textwidth}{(b)}
    \fig{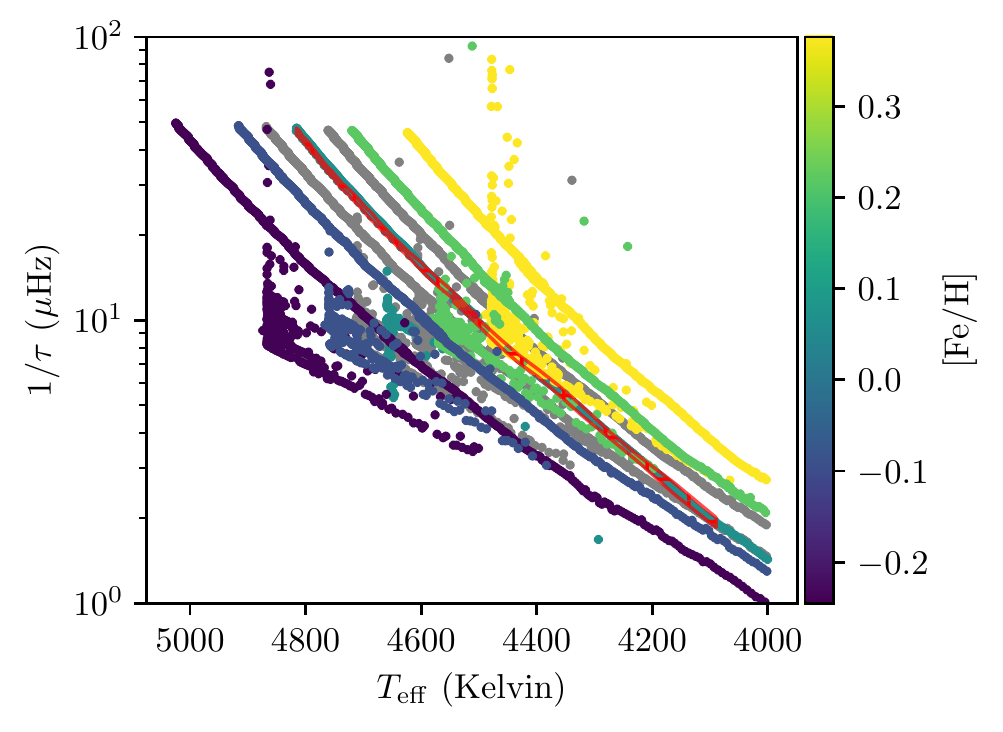}{.45\textwidth}{(c)}
    \caption{Fitted glitch period, $1/\tau$ plotted against effective temperature for stellar models of varying \textbf{(a)} initial mass, \textbf{(b)} $Y_\mathrm{CZ}$, and \textbf{(c)} [Fe/H]. The red crosshatching (appears as a red line here, due to low uncertainty) represents the glitch period uncertainty calculated on the central track. We include the red clump in the figure. The grey lines represent the models from the "fine" grid, which approximate differences in spectroscopic parameters representative of observational errors.}
    \label{fig:per_evol}
\end{figure}

\subsubsection{Glitch Amplitude}

In \cref{fig:amp_evol} we show the fitted glitch amplitude as plotted against effective temperature, varying the stellar mass, helium abundance, and metallicity, respectively. While the glitch signature is ordinarily considered to be a property of the instantaneous surface abundances, these do not substantially change over the course of RGB evolution, particularly since the deep convection zones bring the gravitationally settled helium back to the surface. Consequently, while we show evolutionary tracks coded by surface rather than initial composition, these do not appear to change significantly over each track in our figures.

The amplitude of the glitch signature can be seen to increase with evolution, relative to $\Delta\nu$, as the stars ascend the red giant branch; it therefore appears to serve well as an evolutionary diagnostic (supplementing constraints from $\Delta\nu$ and $\nu_\mathrm{max}$). This result is complementary to that reported in \cite{2014MNRAS.440.1828B}, who reported amplitudes decreasing with $\nu_\text{max}$. While this is true in absolute frequency units, and is also the case with our results, we submit that it is the phases $\epsilon_p$ which carry information about the structure of the star in the mode frequencies, whereas $\Delta\nu$ merely describes the overall size of the mode cavity; as such, it is the dimensionless glitch amplitude (as normalized by $\Delta\nu$) which should be taken to be the fundamentally informative quantity. In each of these cases, we see that the different evolutionary tracks are displaced from each other laterally on this diagram. We attribute this to temperature effects resulting from the differences in stellar mass and composition, rather than differences in the fitted amplitudes: at the same $\log g$, higher stellar masses, higher helium abundances, and lower metallicities each lead to higher  effective temperatures. 

We represent the uncertainty in the glitch amplitude in \cref{fig:amp_evol} by the red crosshatching, shown for one evolutionary track in each figure, in order to avoid visual clutter. Heuristically, the size of this crosshatched area indicates whether the central track can be distinguished from its neighboring tracks. \edit1{For less evolved red giants, the crosshatched regions representing our uncertainties can be seen not to overlap with the adjacent tracks in either the coarse or the fine grids}. Over the course of stellar evolution, the uncertainties in the glitch amplitude increase, so that the tracks become statistically \edit1{indistinguishable} from each other close to the tip of the RGB. Since these fine grids were sampled at a spacing corresponding to typical reported uncertainties in these quantities, this implies that supplementing traditional asteroseismology with additional constraints using the glitch amplitude may not substantially improve the precision of estimates of these parameters (except perhaps \edit1{for the least evolved red giants}), although they may be used as evolutionary constraints in their own right.

\subsubsection{Glitch Period}

We plot in \cref{fig:per_evol} the period of the glitch, $1/\tau$ in $\mu$Hz, against the effective temperature, in a manner similar to  \cref{fig:amp_evol}. The period decreases as the star moves along the red giant branch and increases slightly in the helium-core-burning red clump phase. The color bar indicates how the spectroscopic stellar parameters impact the glitch period. We discuss this only for the red giant phase; the red clump is included in the figure, but there is no obvious relationship between the spectroscopic parameters and the period of the glitch signature \edit1{as the stars relax towards the red clump}. We note that this is tied to the quality of these fits; the stellar structure changes rapidly between each fit during the relaxation period, as the timescale for significant relaxation is smaller than the temporal spacing between each fit. Since our study predominantly focuses on first-ascent red giant stars, we chose not to modify our functional approximation to the glitch signature to ensure the fitting process worked as well on red clump stars and stars in the relaxation period. The outlying points above the curves are examples of these ill-fitted glitch signatures. 

The red giant phase is well-represented by the smooth curves in the figure, which follow the trend of a decreasing glitch period (or increasing acoustic depth) with decreasing temperature. $M$ and $Y$ have a similar qualitative impact on the glitch period; the colored curves show that higher $M$ and $Y$ both result in a lower glitch period at a given temperature, while the opposite is true for [Fe/H] (by the same evolutionary effect as in the previous section). The red crosshatching is again used to visually represent the uncertainty; the glitch period uncertainties appear significantly smaller than those of the glitch amplitude. We thus conclude that the glitch period appears far better constrained than the amplitude by our procedure. Nevertheless, there is clear overlap between the error and the neighboring tracks at low temperatures, just as with the glitch amplitude. We assert that $M$, $Y$ and [Fe/H] measurements are also not substantially improved from use of the glitch period. Nevertheless, given our particular parametrization and fitting procedure, the period is a better choice for relating glitch properties to spectroscopic parameters due to the lower uncertainty.

We observe a separation between the red giant branch and the red clump in \cref{fig:per_evol} as well. In each case, at around 4400-4800K, the red clump is distinguishable from the curves representing the periods of the first-ascent red giant stars. \edit1{This is complementary to the discrimination between RGB and RC stars using $\Delta\nu$ and the radial p-mode phase offset first described by \cite{2012A&A...541A..51K}.}

\subsubsection{Glitch Structure}

\begin{figure}
    \centering
    \includegraphics[width=.45\textwidth]{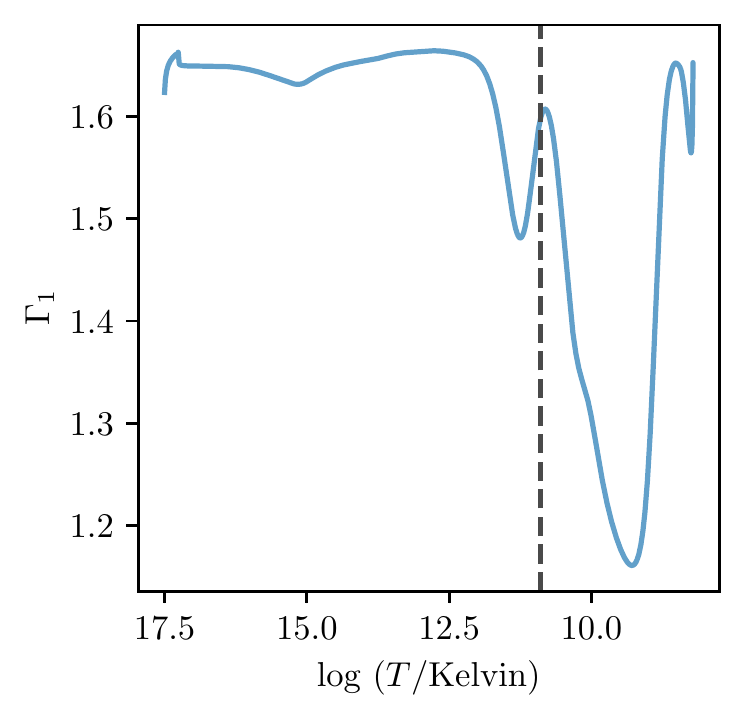}
    \caption{First adiabatic index $\Gone$ of a sample MESA stellar model plotted against effective temperature. The vertical dashed line marks the location of the acoustic depth of the He II glitch, as inferred from the fitted model \cref{eq:parameterisation}.}
    \label{fig:gammaone}
\end{figure}

\begin{figure*}
    \centering
    \gridline{\fig{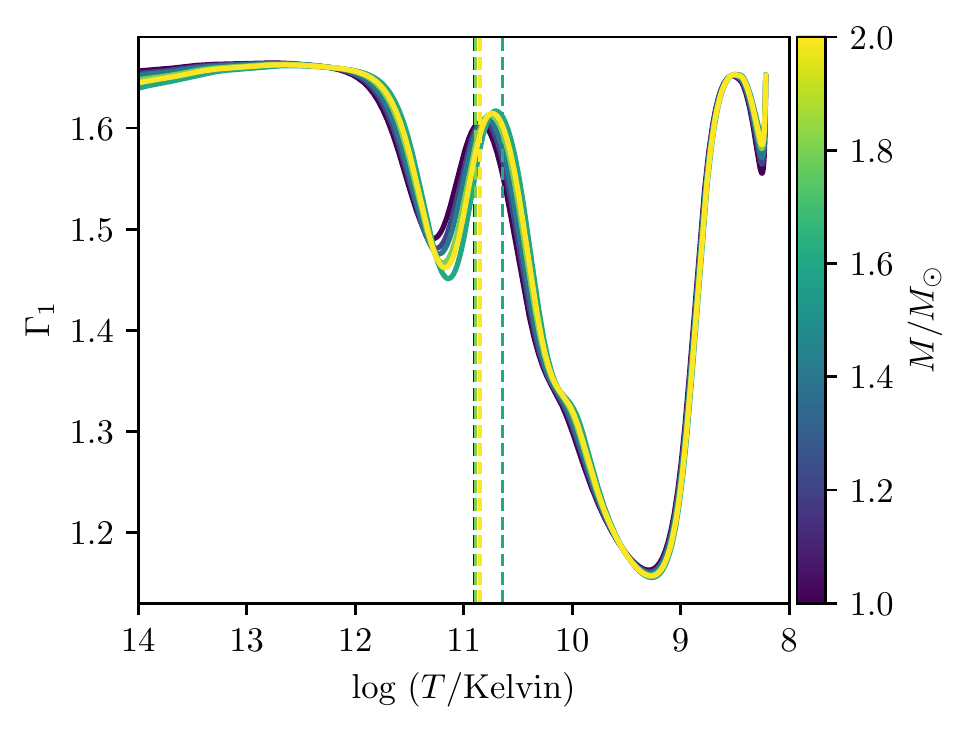}{.4\textwidth}{(a)}
    \fig{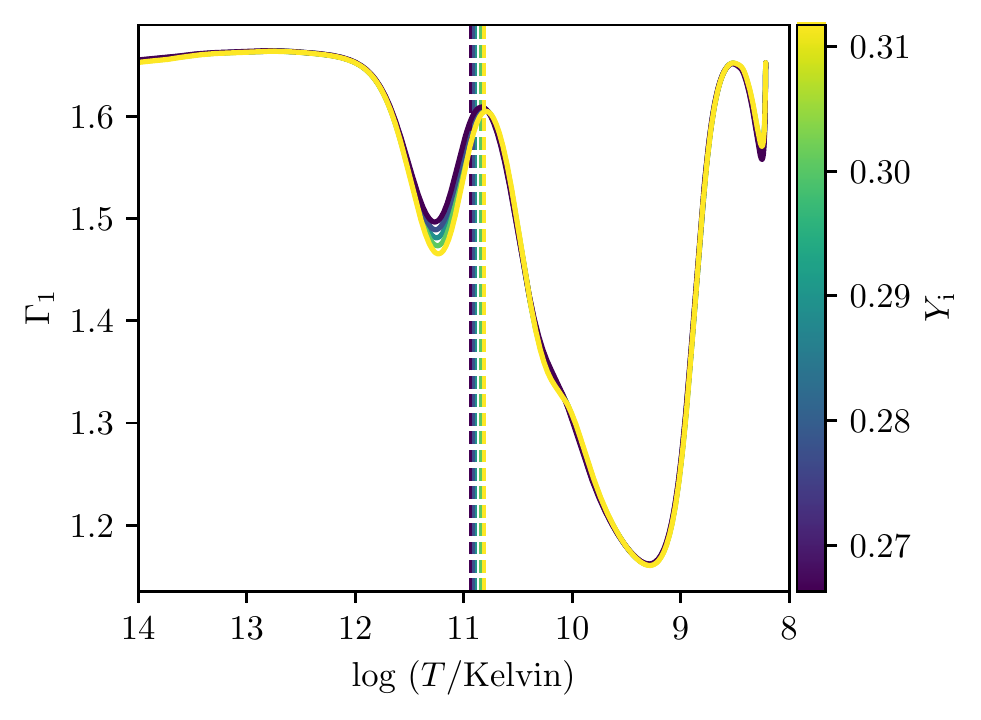}{.4\textwidth}{(b)}}
    \gridline{\fig{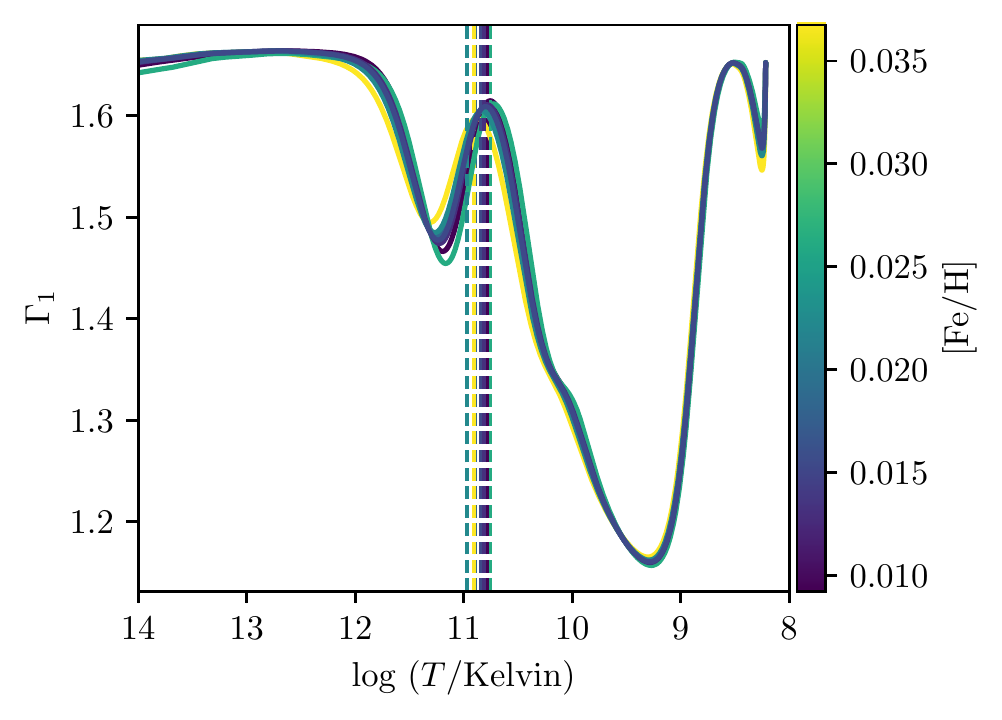}{.4\textwidth}{(c)}
    \fig{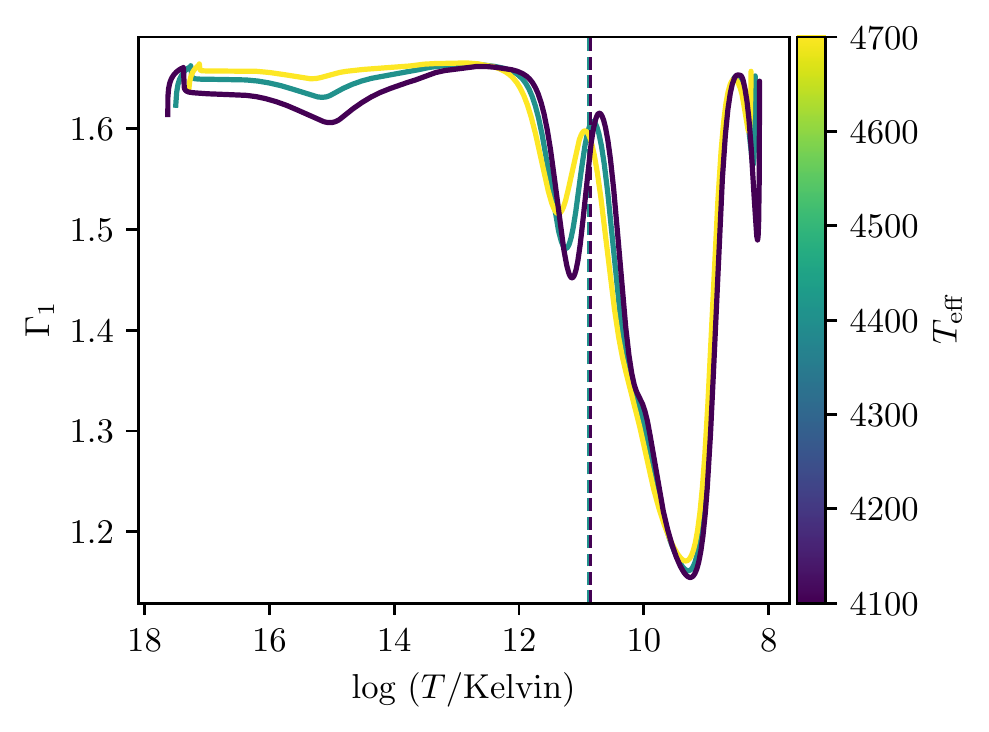}{.4\textwidth}{(d)}}
    \caption{\textbf{(a)} $\Gone$ plotted against effective temperature for stars at a given fixed surface temperature of 4400K. Each curve is colored by the varying $M_\odot$ values. The vertical dashed lines represent the acoustic depth of the He II glitch as computed from our fitted glitch parameters. These are similarly colored by the $M_\odot$ value of the star. \textbf{(b)} The same, but for varying $Y_{cz}$ values. \textbf{(c)} The same, but for varying [Fe/H] values. \textbf{(d)} The same, but for various $T_\mathrm{eff}$ values, or different stages in red giant evolution.}
    \label{fig:GammaOne}
\end{figure*}

We plot the first adiabatic index $\Gamma_1$ against temperature and identify the location of the He II glitch, relating the period to the acoustic depth of the glitch as given in \cref{eq:depth}, computed from the MESA structure file and the fitted period. We see that this localization corresponds to the peak between the two dips in \Gone{} (see \cref{fig:gammaone}), as shown in \citet[for main-sequence stars]{2014ApJ...790..138V} and \citet[for radial modes and p-dominated mixed modes in red giants]{2014MNRAS.440.1828B}. Fits from our procedure, which includes $\pi$-mode frequencies, may thus be interpreted similarly to those done with pure p-modes or p-dominated mixed modes. This continues to be the case when varying the stellar properties across the grid (cf. \cref{fig:GammaOne}), yielding only some minor variations in the localization of the glitch based on the $M$, $Y$, or [Fe/H]. We may treat these minor variations as an estimate of the systematic error incurred in interpreting the glitch period as an acoustic depth, which can be seen to be relatively small. However, we will see that these results are only robust in phases of evolution where the helium glitch lies in the outer half of the star by sound-travel time (cf. \autoref{sec:localisation}). We also examine structural changes caused by evolution on the red giant branch in \cref{fig:GammaOne}. The thermal structure of the star is clearly impacted by evolution, with the acoustic depth of the He II glitch increasing with decreasing $T$. The localization of the He II glitch is consistent with our other findings; at each of the three evolutionary stages plotted, the acoustic depth corresponds to the peak in the $\Gone$ profile between the two depressions.

\section{Dipole Modes}
\label{sec:disc}

We now examine the significance of the inclusion of dipole modes in modelling the glitch. While \cite{2014MNRAS.440.1828B} reported only marginal improvements from supplementing radial modes with quadrupole ones in the glitch-fitting procedure, they were unable to infer dipole p-mode frequencies consistently from dipole mixed modes, and thus reported no improvements from using dipole modes. Claiming to be able to perform this inference, \cite{2020MNRAS.497.1008D} assert that the inclusion of dipole modes substantially modifies the results of the glitch-fitting procedure for red giant stars. Since pure dipole $\pi$-modes are available for our synthetic stars, we are now in a position to validate various aspects of their claims.

\subsection{$\pi$ vs. p-dominated mixed modes}
\label{sec:pi-vs-fit}

The derivation of the underlying pure p-modes associated with an observed set of mixed modes, absent access to the stellar structure, remains an open methodological problem. \edit2{While the frequencies of the most p-dominated mixed modes are a good approximation to those of the pure p-modes for high-luminosity red giants \citep[e.g. in the sample of][]{2021A&A...650A.115D},} \citet{2014MNRAS.440.1828B} found that\edit2{, in general,} attempts to use the p-dominated mixed modes directly to constrain acoustic glitches yielded contradictions between the dipole modes and modes of even degree. Restricting their attention to a single stellar model, \cite{2020MNRAS.497.1008D} proposed and demonstrated one prescription by which these pure modes may be recovered. However, since this work predated the theoretical developments of \cite{ong_semianalytic_2020}, they were unable to critically evaluate the accuracy of this prescription. We are now in a position to better assess the generalizability of these results, which we do by performing a noise-free analysis of a similar kind.

\begin{figure*}[htbp]
    \centering
    \annotate{\includegraphics[width=.45\textwidth]{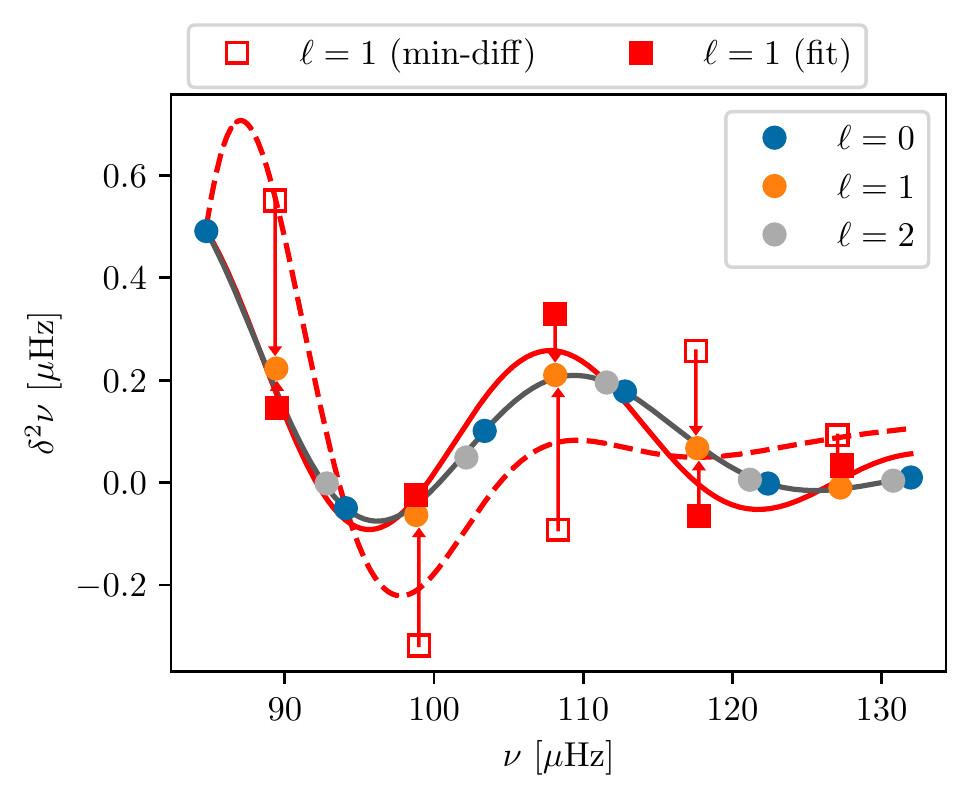}}{\node at (.25, .2){\textbf{(a)}};}
    \annotate{\includegraphics[width=.45\textwidth]{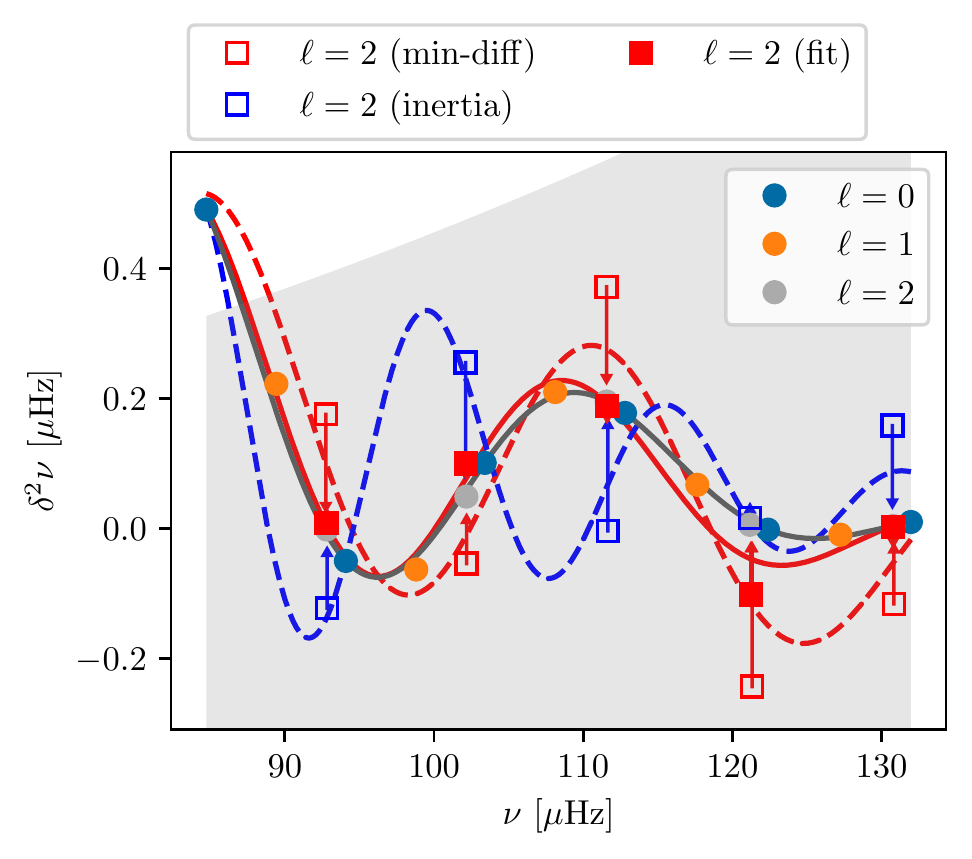}}{\node at (.25, .2){\textbf{(b)}};}
    \annotate{\includegraphics[width=.45\textwidth]{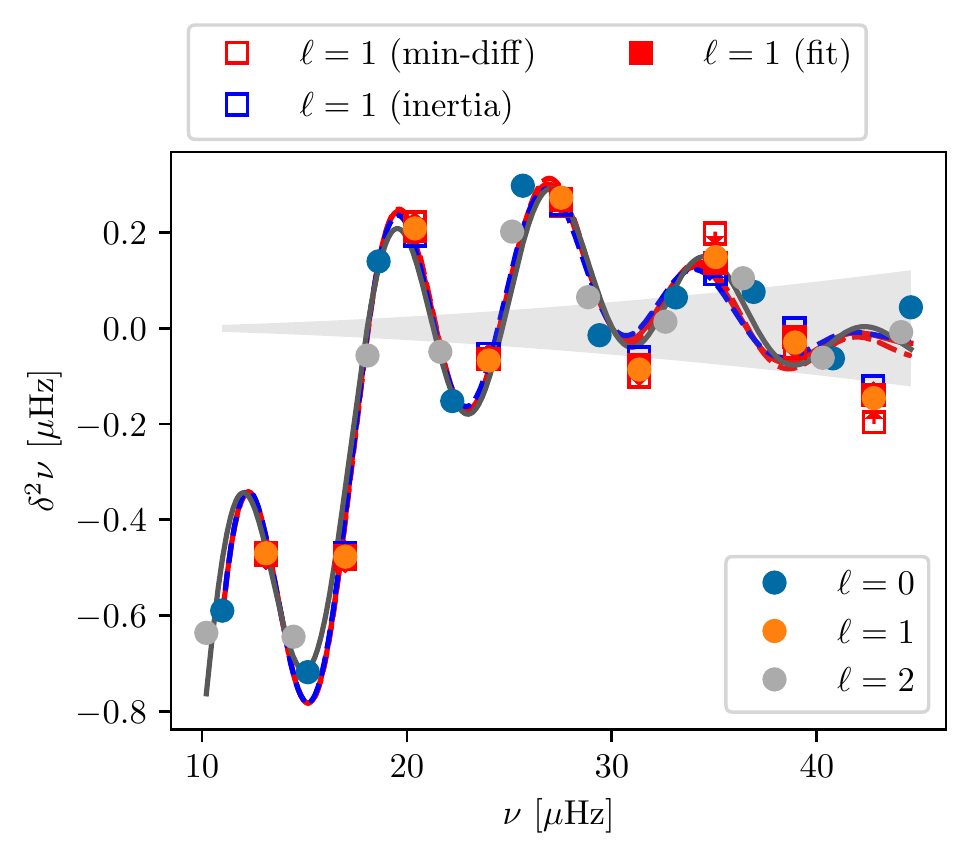}}{\node at (.25, .7){\textbf{(c)}};}
    \caption{Comparison of different constructions for \edit1{nonradial} p-modes and their second differences, \edit2{focusing in particular on dipole modes in \textbf{(a} and \textbf{c)}, and quadrupole modes in \textbf{(b)}. Values in (a) and (b) are from the same stellar model ($\Delta\nu \sim 10\ \mu$Hz)}. Colored circles indicate the second differences of the radial p-modes and nonradial $\pi$-modes computed directly from the stellar structure, while the red squares show the second differences of notional \edit1{nonradial} p-modes as inferred indirectly from \edit1{mixed modes of the same degree} in two different ways, as described in \cite{2020MNRAS.497.1008D}. Red arrows join these indirectly determined quantities to the corresponding values from the $\pi$ modes. The solid black curve shows \cref{eq:parameterisation} as fitted to only the radial modes, while the \edit1{red curves are fitted to the radial modes supplemented with nonradial modes of the corresponding degree (see text for full description). In \textbf{(b)}, the blue squares show second differences of the most p-dominated quadrupole mixed modes, and the blue dashed curve shows a fit to them and the radial p-modes. The gray shaded region shows an implied systematic error interval corresponding \edit3{to the g-mode period spacing, $\delta\nu \sim \nu^2 \Delta\Pi_\ell/2$.}} \edit2{In \textbf{(c)} we show quantities derived from dipole modes recovered from a substantially more evolved red giant model ($\Delta\nu \sim 4\ \mu$Hz) using the same prescriptions, which can be seen to be in much better agreement with the pure $\pi$-modes.}}
    \label{fig:pi-vs-fit}
\end{figure*}

For the purposes of discussion, we first limit our attention to a comparable stellar model (of similar mass and radius; $\Delta\nu \sim 10\ \mu$Hz) to that used in \cite{2020MNRAS.497.1008D}. We compare in \cref{fig:pi-vs-fit}a the second differences in mode frequencies of that model, computed with several different \edit1{prescriptions for the recovery of p-modes from the} dipole mixed modes (cf. their Fig. 2). In particular, filled circles show the second differences of the radial p-modes and nonradial $\pi$-modes of the stellar model, while the square markers indicate quantities inferred from dipole mixed modes. The solid black curve in \cref{fig:pi-vs-fit}a shows our fiducial parameterization, \cref{eq:parameterisation}, as fitted to the second differences of only the radial p-modes (blue circles). The second differences of the nonradial $\pi$-modes (orange and gray circles) lie very close to the fitted curve, despite not contributing to the fit. This is consistent with the analytical properties of acoustic glitches: the inclusion of nonradial modes does not (and indeed should not) materially modify the fitted curve in this noise-free analysis. Conversely, this indicates that we may assess the performance of any observational prescription for deriving p-mode glitch observables from the mixed modes of a stellar model, by way of their consistency with those derived directly from the $\pi$-modes of that stellar model.

The square markers in \cref{fig:pi-vs-fit}a show the second differences of dipole modes calculated using the two prescriptions for finding dipole p-modes from mixed modes considered in \cite{2020MNRAS.497.1008D}: open squares show the results of taking mode frequencies at the local minima of period differences, while filled squares show p-mode frequencies fitted using the asymptotic parameterization of the local period differences described in \cite{cunha_analytical_2019}. As we expect, taking the local minima of period differences (open squares) yields second differences which depart significantly from the glitch profile generated by the \edit1{pure pressure modes: the resulting fit (red dashed curve) is highly inconsistent with them}. The fitting procedure of \cite{cunha_analytical_2019} is intended to remedy this, and indeed can be seen to yield good agreement with the dipole $\pi$-modes at low frequencies. However, this agreement is degraded at higher frequencies, where the g-mode forest is sparser than required to accurately oversample the approximate mode-mixing function $\zeta$ (cf. Ong and Gehan, in prep.). Consequently, a glitch signature fitted against both radial modes and these approximate dipole modes (red \edit1{solid} curve) \edit1{remains visibly different} from that which would be constrained with access to the ground-truth $\pi$-modes.

As such, we expect our characterizations of the helium glitch \edit1{using only pure p-modes}, and the evolutionary dependences we have described above, to differ significantly from those which might be returned when using the approximate dipole p-mode recovery prescription of \cite{cunha_analytical_2019}. This was indeed the case with the analysis of the single stellar model in \cite{2020MNRAS.497.1008D}. \edit1{This divergence was interpreted in that work as a failure of the radial modes to adequately constrain the glitch signature. However, our access to the underlying nonradial $\pi$-modes (as in the preceding discussion) clearly indicates that this not the case. Instead, it is rather the inferred nonradial constraints from mixed modes which are biased, and these differences are almost certainly} a property of this methodological approximation, rather than being of genuine astrophysical significance. \edit1{It is possible} that the putative improvements \edit1{in accuracy (rather than precision)} that \cite{2020MNRAS.497.1008D} suggest to result from the inclusion of dipole modes are perhaps merely fortuitous systematic artifacts of this methodology \edit1{for dipole mixed modes}.

\edit1{In addition to dipole modes, we also examine the accuracy of various approximations for recovering p-mode glitches from the quadrupole modes in \cref{fig:pi-vs-fit}b. We show with blue markers in \cref{fig:pi-vs-fit}b the second differences of the frequencies of the most p-dominated quadrupole mixed modes, which are often used to approximate those of the underlying quadrupole p-modes. The blue dashed curve shows \cref{eq:parameterisation} as fitted to them in combination with the radial modes. We see that it is very significantly discrepant from the actual glitch signature implied by the radial p- and nonradial $\pi$-modes. As such, we conclude that this commonly-used approximation may not be sufficiently accurate for the purposes of constraining the acoustic glitch. Since the mode coupling for dipole modes is stronger, and the period spacings are larger, this means that the use of a similar approximation for the dipole modes is even less appropriate than for quadrupole modes.}

\edit1{We note that these differences emerge from a noise-free analysis, whereas in principle, this bias could potentially be reduced by downweighting the quadrupole modes in the fit (i.e. artificially inflating the associated measurement errors) to account for the fact that the frequencies of these minimal-inertia mixed modes necessarily deviate from those of the underlying p-modes. A priori, this deviation is at most $\delta\nu_2 \sim \nu^2 \Delta\Pi_2$; we use it informally as an implied estimate of the systematic error associateed with this approximation. We show the size of this systematic error with the gray shaded region (centered at 0) in \cref{fig:pi-vs-fit}b. We see that it is so large as to be comparable to the amplitude of the glitch signature itself. Accordingly, were this downweighting to occur, the quadrupole modes would have essentially no meaningful constraining power on the properties of the acoustic glitch. Similar arguments should also apply to the minimum-period-difference technique for the dipole mixed modes.}

\edit1{Next, we consider the above approaches to deriving pure p-modes from mixed modes, applied to the quadrupole mixed modes of the same model. These are shown using red symbols in \cref{fig:pi-vs-fit}b, with the same meaning as in \cref{fig:pi-vs-fit}a. We see now that the use of the asymptotic parameterisation of \citet{cunha_analytical_2019} now yields results (red filled squares and solid curve) that are in very close agreement with those arising from the quadrupole $\pi$-modes (gray circles). To our knowledge, the application of these methods to quadrupole modes has not been well investigated, as observations of these modes have not been reported. This paucity of observations is caused by the difficulty of exciting more than the one p-dominated quadrupole mixed mode per radial order to observable amplitudes.}

\edit2{Finally, we examine the limiting behaviour of these constructions in the regime of high-luminosity RGB stars, where the coupling between the p- and g-mode cavities becomes extremely weak, and the density of g-modes becomes extremely high. For such red giants, the coupling between the two mode cavities is typically neglected, and the p-dominated dipole mixed modes are treated as p-modes for glitch analysis \cite[e.g.][]{2021A&A...650A.115D}. We show in \cref{fig:pi-vs-fit}c these constructions applied for the recovery of dipole modes in a substantially more evolved RGB stellar model ($\Delta\nu \sim 4\ \mu$Hz). In this regime of evolution the most p-dominated mixed mode frequencies are assumed to be good approximations to those of the underlying pure p-modes, and indeed the second differences of the inferred p-mode frequencies can be seen to be in very close agreement with the $\pi$-modes of this stellar model. The systematic errors incurred from neglecting mode coupling (gray shaded region) are also considerably smaller in this regime, even for dipole modes.}

\subsection{Influence of dipole modes on fitted properties}

\begin{figure}
    \centering
    \fig{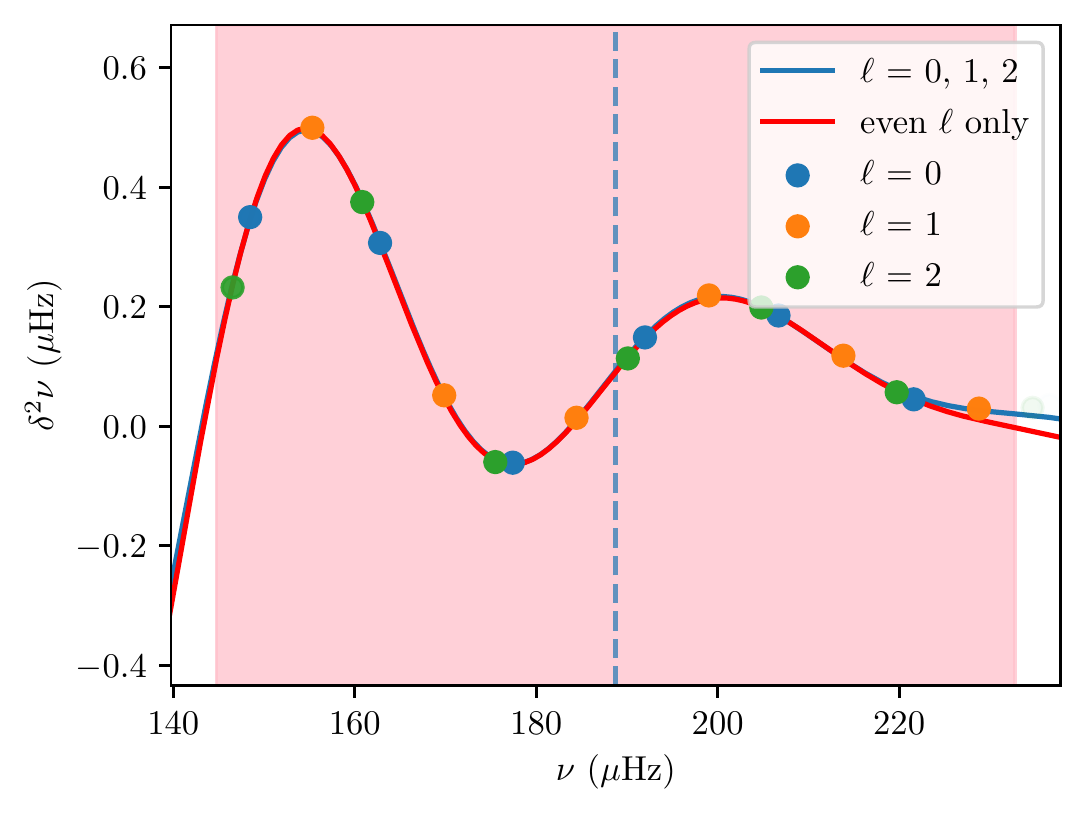}{.45\textwidth}{(a)}
    \fig{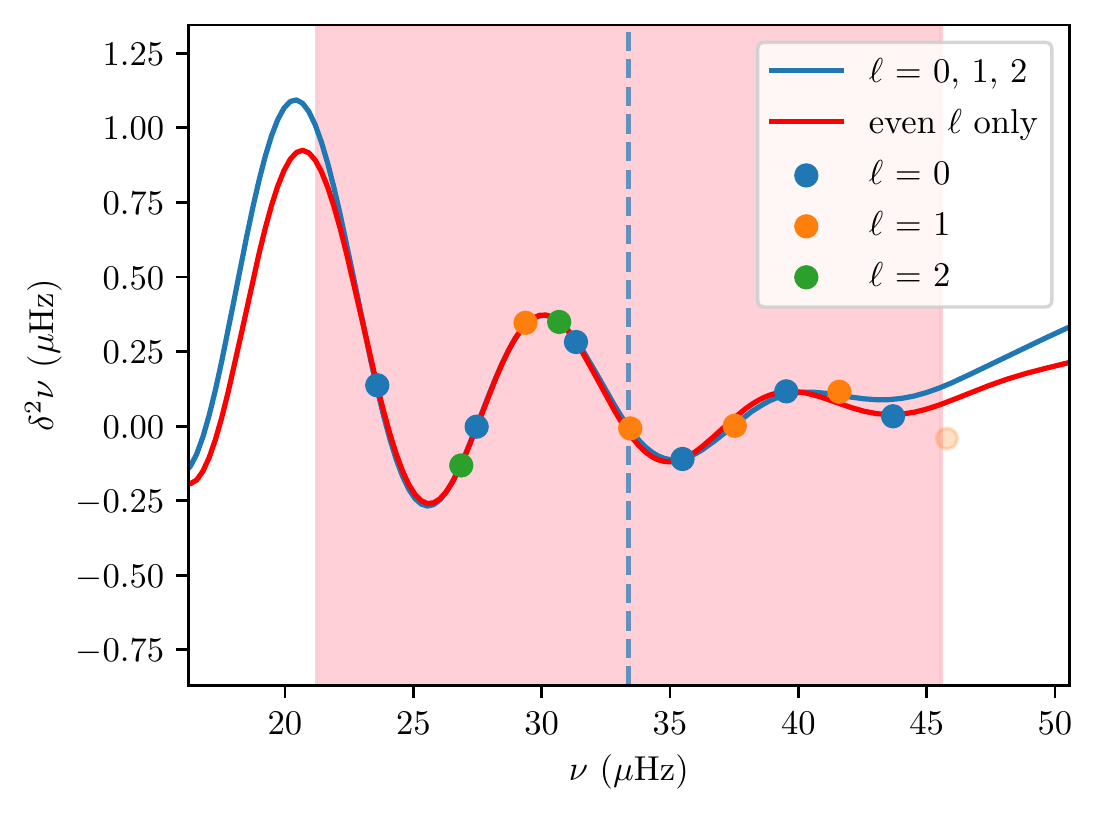}{.45\textwidth}{(b)}
    % \fig{glitch_track.pdf}{.45\textwidth}{(c)}
    \caption{\textbf{(a)} Second differences of a sample glitch signature for an early RGB star, fitted via two different methods. Colored circles indicate the $\ell = 0, 1, 2$ modes. The solid blue line is a fit using $\ell = 0, 1, 2$ modes, while the red line is a fit without $\ell = 1$ modes. The dashed vertical line represents the location of \numax. \textbf{(b)} The same, but for a more evolved RGB star.}
    \label{fig:glitchtrack}
\end{figure}

\edit1{\cref{fig:pi-vs-fit} shows that in some cases, the inclusion of pure dipole p-modes does not change results obtained with only pure p-modes of even degree. A priori we expect this to not necessarily always be the case, and we investigate such differences in this section.} We compare in \cref{fig:glitchtrack}a and b the quality of fit with and without the use of dipole modes, at two different ages along the same evolutionary track. The two fitted curves are nearly identical around \numax, but trend away from each other at high and low frequencies. This result is more pronounced for the fits done at a later evolutionary stage, as is visible in \cref{fig:glitchtrack}b.  We conclude that, on a model-by-model basis, the inclusion of dipole modes in the fitting procedure results in a fit largely consistent with one produced using only even degree modes.

\begin{figure}[htbp]
    \centering
    \includegraphics[width=.45\textwidth]{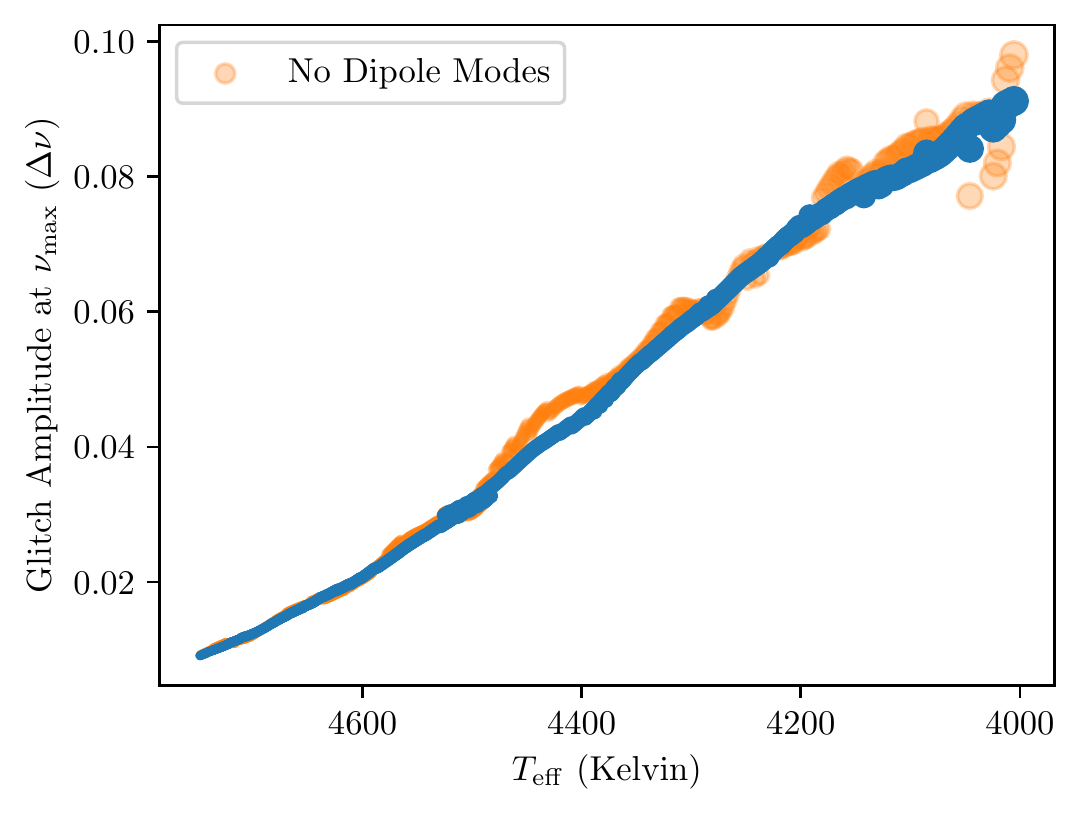}
    \caption{Glitch amplitude at $\numax$ plotted against effective temperature across RGB evolution for a MESA-generated stellar model. The blue points represent amplitudes fitted with the use of $\ell = 1$ modes, while the orange points represent amplitudes fitted without $\ell = 1$ modes. The luminosity bump is visible at around 4500K.}
    \label{fig:diptrack}
\end{figure}

We examine in more detail in \cref{fig:diptrack} how these differences change over the course of stellar evolution. The fitted amplitudes with and without dipole modes can be seen to differ slightly from each other: amplitudes fitted with dipole modes evolve smoothly, while those fitted without them exhibit small oscillatory excursions. These excursions increase in magnitude (i.e. fits without dipole modes become increasingly inaccurate) for more evolved models at lower temperatures, coinciding with where \cite{2014MNRAS.440.1828B} find that even-degree modes alone cease to robustly constrain the fitted amplitudes and depths. We conclude from this that the inclusion of dipole p-modes, were they available, would in general significantly improve the robustness of the glitch modelling procedure in red giants. \edit1{However, given the methodological issues involved with inferring p-mode frequencies from mixed modes \edit2{that we have considered in \autoref{sec:pi-vs-fit}}, we feel it important to qualify that such constraints on dipole p-modes should only be introduced where their availability is considered reliable. We do not believe this to be the case with present techniques for the analysis of dipole mixed modes}

\subsection{Localization of the helium glitch}
\label{sec:localisation}

When fitting for the acoustic depth of the glitch, we have assumed (in keeping with the usual practice for main-sequence stars) that the glitch is localized in the outer half of the star, by sound travel time. This assumption is enforced by the hard cutoff used in our initial parameter sweep --- see \cref{eq:outer}. However, this assumption may not necessarily hold in the most evolved red giants, where the convective envelope becomes extremely distended. In practical terms, this assumption is also motivated by the fact that modes of each degree sample the glitch signature at roughly intervals of $\Delta\nu$, and, since $\Delta\nu \sim 1/2T$ ($T$ being the acoustic radius), would therefore by themselves have difficulty distinguishing between localizations of the glitch at its correct location $\tau$, or at the alias $T-\tau$ (i.e. so that the sinusoidal frequency provides an acoustic radius rather than depth). In principle, the use of modes of different $l$ (and in particular different parity of $l$) should significantly alleviate this degeneracy.

\begin{figure}[htbp]
    \centering
    \includegraphics[width=.45\textwidth]{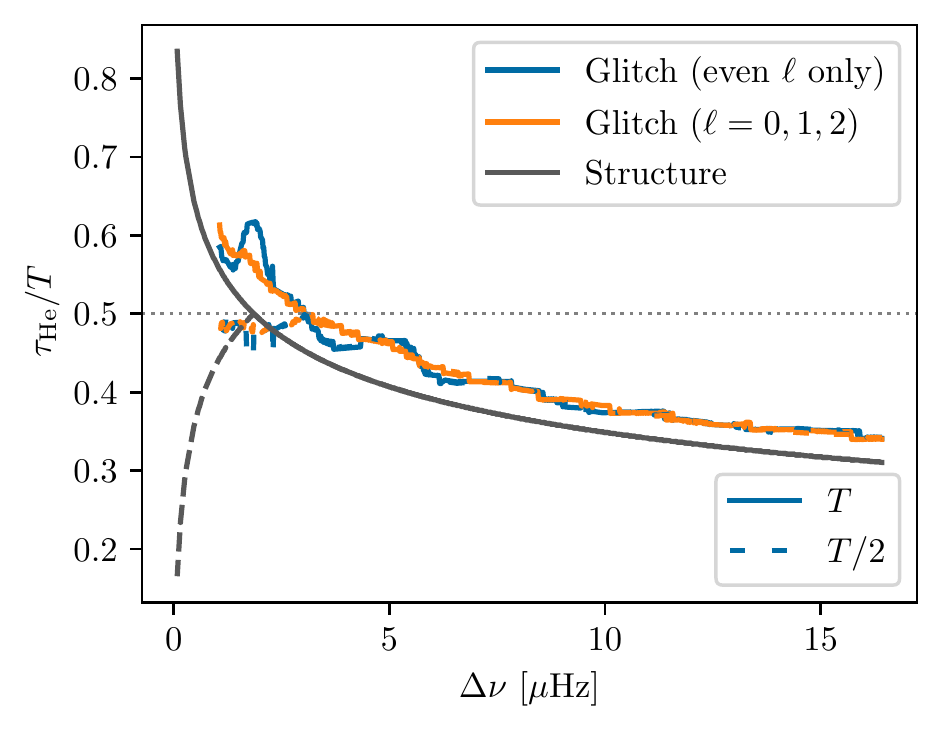}
    \caption{The fractional acoustic depth of the helium glitch, $\tau_\mathrm{He}$, along an evolutionary track of solar composition at $1.2 M_\odot$, shown as a function of $\Delta\nu$ (i.e. with evolution going from right to left), as determined from different variations on our method. The locations implied by two different glitch fits (with and without dipole modes) are shown with the blue and orange curves, while the gray curve shows the ``true'' location of the glitch as determined directly from the stellar structure. The horizontal dotted line shows the sampling rate of $\Delta\nu$, while the gray dashed curve shows the alias of the gray curve around this sampling rate.}
    \label{fig:taualias}
\end{figure}

We show with the solid gray curve in \cref{fig:taualias} the notional location of the glitch, as determined from stellar models by directly evaluating the acoustic depth at which the adiabatic index $\Gamma_1$ attains a local maximum (as shown in \cref{fig:gammaone}). The gray dashed curve shows the same quantity, but aliased against a \edit1{notional} repetition rate of $\Delta\nu$ to yield values always less than $T/2$ (marked out with the horizontal dotted line). Correspondingly, the colored curves indicate the locations of the glitch implied by the fitted sinusoidal frequency, either using the hard cutoff in our parameter sweep (as in \cref{eq:outer}, shown with dashed curves), or with the parameter sweep widened to encompass potential aliases (solid curves).

As was the case with the fitted amplitudes, the fitted acoustic depths of the glitch can be seen to exhibit oscillatory variations over the course of stellar evolution when only even-degree modes are used, compared to when the dipole modes are included in the constraint. However, the overall qualitative characteristics in both cases remain similar. Again, these results are different from the findings of \cite{2020MNRAS.497.1008D}, for reasons that we have already examined.

We see also that the constraint imposed by \cref{eq:outer} remains largely \edit1{a valid description of the true localisation of the glitch} until very far up the red giant branch, past the luminosity bump ($\Delta\nu \lesssim 3\ \mu$Hz): only there do the analyses with and without the cutoff of \cref{eq:outer} diverge. While relaxing this constraint does permit our methodology to be applied to more evolved stars, we see that this is only efficacious within a narrow range of evolutionary states: \edit2{far beyond the luminosity bump}, the true acoustic depth of the enhancement in $\Gamma_1$ increases sharply. \edit1{However, in this regime, our fitting procedure does not usually succeed in returning glitch parameters}. The scaling between $\nu_\text{max}$ and $\Delta\nu$ is such that the radial order $n_p$ of modes near $\nu_\text{max}$ decreases with evolution, yielding fewer modes within our chosen frequency window centered on $\Delta\nu$ than would be available for less evolved ones \citep{2014ApJ...788L..10S}. For these highly evolved stars, the number of available modes has decreased to such an extent that the number of free parameters in our parameterisation is larger than can be constrained by the data. While using dipole modes does increase the number of available second differences, this phase of evolution is sufficiently rapid that their use here does not substantially assist in resolving this issue.

\section{Observational Systematics}
\label{sec:obs}

In order to assess the usefulness of our parameterization in analyzing real data, a handful of stars were selected from light curves produced using the KEPSEISMIC pipeline \citep{kepseismic1,kepseismic2} and passed through our glitch fitting process. \edit3{Our interest in this section is methodological, rather than astrophysical; as such, we} selected bright stars ($< 8.5$ Kepler magnitude) in the first-ascent red giant phase. \edit3{Our selection of bright stars is not intended to be representative of red giants collectively, but rather to avoid interactions between systematic issues caused by having a larger proportion of intrinsic noise in the data}. We then sorted the stars by the length of the time series and took those with the longest time series, since our analysis relies on slicing said time series into numerous different lengths. Stars without well-defined $\ell$ = 0 and $\ell$ = 2 mode ridges were excluded from the study. Peakbagging was done using the PBJam code \citep{2021AJ....161...62N}. An additional manual step followed the \edit1{initial} generation of the acoustic modes; we vetted PBJam’s ridge identification of the radial orders, and thus \edit1{eliminated from our sample those} stars with low SNR and incorrect identifications. From these considerations, we settled on 9 stars, from which we used the 55-day light curve filter. These stars are given in \cref{table:1}. \edit2{In addition to their posterior median $\Delta\nu$, as determined by PBJam's ``asymptotic peakbagging'' routine, we also} \edit1{report their \edit2{single-mode height-to-background ratios (which for brevity we will refer to as a signal-to-noise ratio, or SNR)} as defined by the maximum height-to-background ratio of any radial mode fitted by PBJam, which normalises mode heights with respect to the coloured-noise components of the power spectrum}. \edit2{We note that this construction is defined with respect to individual modes, rather than to the power spectrum as a whole. As such, it is considerably more sensitive to local stochasticity in the power spectrum than the usual definition of the height-to-background ratio, which involves smoothing out the power spectrum to yield an averaged amplitude \cite[e.g.][]{2008ApJ...682.1370K,2011ApJ...743..143H,2012A&A...537A..30M}. This local sensitivity is necessary, since our primary concern here is the robustness of measurements of the glitch properties (derived from individual mode frequencies), rather than of the amplitudes of the seismic power excess per se}. \edit3{We report this SNR to characterize the information content of the power spectrum as a point of comparison for the computed glitch parameters of different stars. In the actual fitting procedure, we use 8 radial orders around $\nu_\text{max}$ as is standard with PBJam}. The rest of the pipeline is fully automated, taking the generated frequencies and fitting them, just as with the artificial data. The same model parameterization given in \autoref{eq:parameterisation} was used. The only notable difference was that dipole modes were not \edit1{returned by pbjam}, and so were not used in the fitting process. 

\edit2{In the following sections, we consider the response of the fitted glitch properties to simulated progressive degradation of observing conditions. Accordingly, we also show in \autoref{table:1} the median SNR for each star at the end of each of these simulated degradation exercises. In particular, in \autoref{sec:duration} we consider how the glitch parameters are modified when the observation window is truncated; in \autoref{table:1} we therefore report the median SNR for each star over all truncated windows of the shortest duration used in the exercise. Likewise, in \autoref{sec:injection} we consider how they change when noise is injected into the light curve; we therefore report the median SNR taken over all realisations of the largest amount of injected Gaussan noise used in the exercise. In both cases, we then take the ratios of these median maximally degraded SNRs, compared to the ones returned from the unmodified Kepler data, to heuristically indicate the change in the information content of the seismic signal within each Kepler light curve, under the action of each kind of degradation. In our following discussion we will refer to these as the ``truncation" and ``degradation" ratios, respectively.}

\begin{table}
\centering
\caption{KIC Numbers of selected stars whose stellar light curves were produced from the KEPSEISMIC pipeline along with \edit2{various measures of their SNR (defined in the main text)}.}
\scriptsize{
\tabularnewline
\begin{tabular}{ cc|ccc|cc } 
    \toprule
    \multirow{2}{*}{KIC Number} & \multirow{2}{*}{$\Delta\nu/\mu$Hz} & \multicolumn{3}{c}{SNR} & \multicolumn{2}{c}{Ratio} \\
    & &  \scriptsize{Raw} & \scriptsize{Truncated} & \scriptsize{Degraded} & \scriptsize{Truncation} & \scriptsize{Degradation} \\ 
    \midrule
    7286856 & 14.7 & 61.436 & 20.000  & 3.195  & 0.33 & 0.05 \\
    8631401 & 11.5 & 35.256 & 16.858  & 3.586  & 0.48 & 0.10 \\
    6144777 & 11.0 & 81.685 & 17.452  & 8.104  & 0.21 & 0.10 \\
    11352446 & 7.7 & 45.498 & 9.766   & 4.874  & 0.21 & 0.11 \\
    11618103 & 9.4 & 52.735 & 16.472  & 7.634  & 0.31 & 0.14 \\
    8328178 &  8.6 & 64.325 & 15.113  & 16.365 & 0.24 & 0.25 \\
    5790837 &  4.7 & 24.415 & 7.593   & 7.744  & 0.31 & 0.32 \\
    7668613 &  4.3 & 31.926 & 3.750   & 10.990 & 0.12 & 0.34 \\
    7944142 &  7.1 & 33.809 & 14.470  & 12.398 & 0.43 & 0.37 \\

    \bottomrule
\end{tabular}
}
\label{table:1}
\end{table}

\subsection{Dependence of Amplitude Uncertainty on Length of Time Series}
\label{sec:duration}

\edit2{We first examine how the uncertainties in the fitted glitch amplitudes change as the available duration of the time series is decreased. For this exercise, we consider window lengths of varying durations (from 27 days up to 2 years, in increments of 27 days). For each duration, we prepare 10 randomly chosen slices of the Kepler time series for each star. We then pass each slice through the fitting pipeline described in \autoref{sec:methods}. The final uncertainties in the glitch amplitudes (for this exercise) are then found as the standard deviation of the fitted value across all windows of the same duration.}

\begin{figure}
    \centering
    \includegraphics[width=9cm]{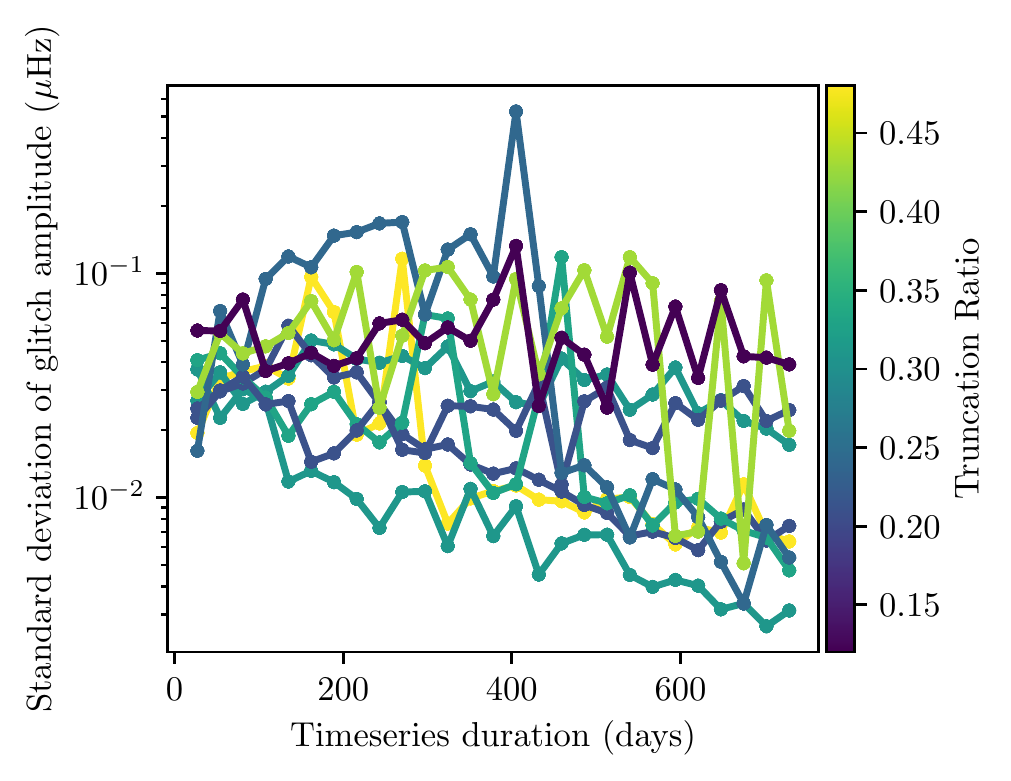}
    \caption{The standard deviation in the glitch amplitude of 9 stars whose light curves were produced using the KEPSEISMIC pipeline, plotted against the duration of the time series. The points represent each standard deviation calculation, and the lines connect a given star for visual clarity. Lines are colored based on the \edit2{star's truncation SNR ratio}.}
    \label{fig:duration}
\end{figure}

\begin{figure}
    \centering
    \includegraphics[width=9cm]{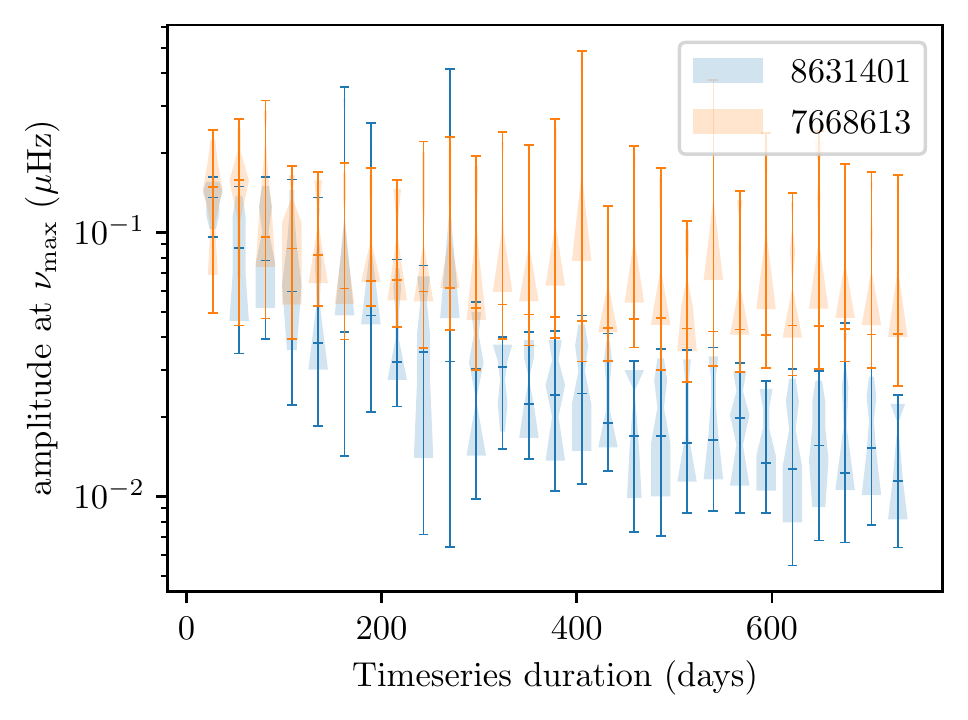}
    \caption{\edit3{Violin plot of the glitch amplitudes of two selected stars plotted against timeseries duration.} The blue lines represent amplitudes of the star in our sample \edit2{whose SNRs were least affected by truncation}, and the shaded regions represent the probability distributions around each median. The orange lines are of the same nature, but for the star in our sample \edit2{most affected by truncation}.}
    \label{fig:violin}
\end{figure}

\cref{fig:duration} illustrates that this uncertainty in the amplitudes marginally decreases as the length of the time series increases. The numerous fluctuations for each given star (colored by \edit2{truncation SNR ratio}) indicate that this relationship is not \edit1{deterministically} linear. \edit2{We examine this more closely in} \cref{fig:violin}, which shows how differences in reported amplitudes depend on duration, \edit2{for two stars with different truncation SNR ratios}. \edit2{Despite both of these stars having very similar intrinsic SNRs in their unmodified \textit{Kepler} data, they can be seen to respond very differently to this simulated progressive degradation.}

\edit2{In both cases, we see that the fitted amplitudes are fairly inaccurate for the shortest-duration windows, and converge towards limiting values as the durations of the windows increase. In both cases, we also see that the fitted values of the glitch amplitude appear overestimated on average for the most truncated windows; this appears to be a common feature of the sample that we have considered. For KIC 8631401  (where the peakbagging was least affected by truncation in our sample), we find that the uncertainties in the fitted glitch amplitude are consistently smaller, and the distribution of fitted amplitudes are more centered around the median, than appears to be the case for KIC 7668613 (whose data were the most sensitive to truncation of our sample). Surprisingly, despite the truncation ratios indicating that KIC 8631401 should be least affected by truncation, we find that its fitted glitch amplitude changes far more, as the amount of available data increases. One possible reason for this is that KIC 7668613 is the most evolved star (smallest $\Delta\nu$) of our bright sample. Prima facie, we should therefore expect its glitch amplitude to be the smallest in absolute frequency units, from the homology scaling considerations that we have discussed in our modelling exercise. However, we can see that the fitted amplitudes are uniformly larger than those of KIC 8631401, which is considerably less evolved. Thus, one likely explanation for this is that even a 2 year temporal baseline, which is the longest that we have considered in this exercise, remains insufficient to provide the data quality required to adequately constrain the properties of the helium glitch in KIC 7668613.}

\subsection{Dependence of Amplitude Uncertainty on background noise}
\label{sec:injection}

\begin{figure}
    \centering
    \includegraphics[width=9cm]{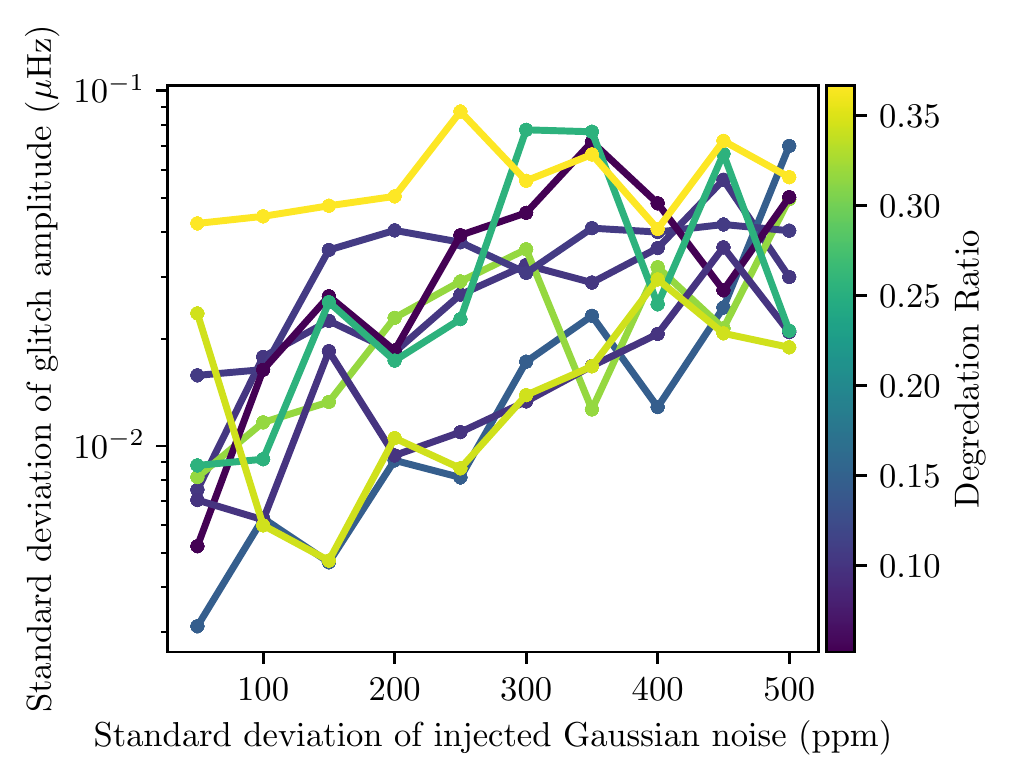}
    \caption{The standard deviation in the glitch amplitude of 9 stars whose light curves were produced using the KEPSEISMIC pipeline, plotted against the standard deviation of the added Gaussian noise. The points represent each standard deviation calculation, and the lines connect a given star for visual clarity. Lines are colored based on the star's \edit2{degradation SNR ratio}.}
    \label{fig:snr}
\end{figure}

Next, we examine the relationship between the uncertainty in the fitted amplitude and \edit2{the statistical properties of the non-oscillatory components in the input time series}. Gaussian noise was injected into each time series to simulate progressive degradation of the intrinsic photometric noise of each target. \edit2{Given that the duration of the original \textit{Kepler} time series for each star was slightly different, we chose an arbitrary fixed length of 180 days to standardise the total amount of information available between stars.} \edit1{Ten different input standard deviations in the range of 50-500\,ppm were used to generate temporally uncorrelated noise, which was then injected into the flux of each time series prior to mode identification via PBJam.} We considered 10 independent realizations of this white noise for each simulated degraded SNR. This gave us 100 uniquely-altered light curves for each star, for which the mode identification and fitting process was redone \edit1{ab initio as described} in \autoref{sec:duration}. \edit1{In \cref{fig:snr}, we show how the uncertainties in the glitch amplitude depend on the standard deviation of the injected photometric noise. As expected, the injection of noise increases the uncertainty in the resulting \edit2{fitted glitch} amplitudes. We see that the \edit2{degradation ratio} does not strongly correlate with \edit2{the overall normalisation of} the amplitude uncertainties in a deterministic fashion}. \edit2{However, the degradation ratio can be seen to correlate well with the overall slope of the relation between the scatter in the fitted amplitudes, and the amount of injected Gaussian noise. Moreover, we find that the errors in the glitch amplitude} have a clearer monotonic dependence on the injected noise than \edit2{they do on the duration of the associated time series}.

\subsection{Discussion}

\edit2{We have examined how our ability to recover glitch parameters for red giant stars is affected by the time-domain properties of the input data, specifically the duration and noisiness of the associated time series, by way of simulating the progressive degradation of observing conditions. We have used the ratios of single-mode HBRs to quantify the amount by which the seismic component of the input data is modified under both kinds of degradation. We find that stars with similar intrinsic \textit{Kepler} SNR respond differently in these exercises.}

\edit2{Informally speaking, we should obtain the most drastic changes to the power spectrum (and so the smallest degradation ratios) for data sets which are most informative, while data sets which are already noise-dominated should change the least (and so yield the largest degradation ratios). Although increasing the amount of white noise in the input data can be directly seen to worsen the quality of the fitted glitch parameters, the dependence on the length of the input time series is less clear; correspondingly, the interpretation of the truncation ratios is somewhat less straightforward. In principle, we should expect that the truncation ratio should depend on $\Delta\nu$, to the extent that longer temporal baselines are required to resolve smaller frequency spacings. However, we find at best only a weak positive correlation with the values of $\Delta\nu$ that we report in \cref{table:1}. While the truncation ratio does appear to parameterise the rate at which the fitted properties approach their limiting values as the length of the available time series is increased (e.g. \cref{fig:violin}), the relationship between this truncation ratio and the photometric properties of the time series (i.e. the raw single-mode HBR) is also unclear.}

\edit2{As far as seismic characterisation of glitch signatures is concerned, the relative importance of photometric stability of the time-series data, compared to the duration of the time series, appears to depend on the regime of evolution of the intended targets. A long time series is fundamentally necessary to obtain a reliable amplitude for high-luminosity red giants. However, for less evolved stars, this improvement in the fitted parameters saturates as the duration of the available data increases. On the other hand, a large amount of photometric noise makes such a determination impossible, regardless of the length of the time series.}

\edit3{This tradeoff may have implications for the design of future asteroseismic surveys. The results of our modelling exercise, e.g. \cref{fig:amp_evol}, are suggestive of evolutionary threshholds for both the minimal duration of time series, as well as for the maximal permissible amount of photometric noise, required for reliable extraction of the properties of the glitch. For example, for the most evolved stellar models on our computational grids, the fitted glitch amplitudes are of order $\sim 0.1\ \mu$Hz; heuristically, \cref{fig:duration} suggests that time series longer than 300 days, and \cref{fig:snr} that no more than 300 ppm of photometric noise per exposure, will be correspondingly required to reliably characterise the glitch for stars in similar phases of evolution. More detailed diagnostics will naturally depend on the precise instrumental characteristics and observing strategy for the survey under consideration, which we believe to lie beyond the scope of this work.}

% With a more detailed diagnosis of glitch amplitude uncertainty in this section, we highlight the significance of a star having a high intrinsic SNR. {\color{red} this will have to be rewritten. Also, how much overlap do we want?}\cref{fig:violin} shows KIC 8631401 and KIC 7668613 both have high errors and inconsistent amplitude measurements at low time series durations. We note that KIC 7668613 is equally noisy at higher durations, while the error shrinks drastically and the amplitude reaches a more consistent value for KIC 8631401. We conclude that, under observational constraints,  

% We note that in \cref{fig:snr} the intrinsic SNR does not appear to correlate with the reported uncertainty in the glitch amplitude. However, stars with higher SNR show dependences on the injection of additional noise that are more clearly \edit1{monotonic}, while the lower SNR stars show more fluctuations, particularly when the amount of injected noise is high (400-500\,ppm).

\section{Conclusion}
\label{sec:conc}

In this paper, we have developed a fitting procedure for the He II glitch of first-ascent red giant \edit2{stellar models}. We then use it to assess the potential use of glitch parameters to constrain stellar properties, as well as methodological systematics associated with the inclusion or omission of dipole modes in improving the fits of red giant He II glitches.

Using a grid of evolutionary models, we identify relationships between the glitch parameters and other spectroscopic parameters. Under reasonable assumptions about the observational uncertainties in the mode frequencies, we find that \edit1{the inferred uncertainties} in the glitch amplitude and period \edit2{(derived via the mode frequencies generated with the stellar models)} increase with evolution as well. \edit1{Thus,} measurements of $Y$, $M$, and [Fe/H] \edit1{do not appear to be} substantially improved with the constraint of the glitch amplitude or period, except perhaps for \edit1{the least evolved red giants}. In contrast, we find that the period of the glitch signature can differentiate the red clump and red giant stages, in conjunction with measurements of the effective temperature. \edit2{\citet{2021A&A...650A.115D} distinguish the red clump and red giant evolutionary stages on the basis of the fitted phase parameter of the glitch, rather than period. However, by construction, a glitch signature fitted to the second differences of the mode frequencies will have a different phase parameter from one fitted directly to phased frequencies, particularly if the chosen parameterisation of the amplitude function is different. Thus, the phases we obtain are incommensurate with those in \citet{2021A&A...650A.115D}.} The computation of $\tau$ from our fitted period also conforms to earlier results \citep{2014ApJ...790..138V, 2014MNRAS.440.1828B} that the He II glitch’s localization corresponds to the peak between the two depressions in the $\Gone$ diagram. This statement appears to hold for evolutionary models of varying $Y$, $M$, and [Fe/H].

Following \citet{2020MNRAS.497.1008D}, we investigate the effects of including dipole modes using the $\pi$-mode isolation scheme of \cite{ong_semianalytic_2020}. We find that the use of dipole modes does not significantly alter individual fits, though they become important when fitting entire evolutionary tracks \edit2{approximately at the luminosity bump and beyond}. Our results highlight shortcomings in present methods in inferring dipole \edit1{and quadrupole} p-modes from the available mixed modes.

Finally, we tested our fitting procedure on \emph{Kepler} light curves in order to benchmark the study and understand the limitations imposed by the frequency errors of real data. We found that our fitting procedure applied itself smoothly to these data. \edit2{We then explored how the fitted properties of the glitch are modified under different kinds of degradation of observing conditions.}

In conclusion, we have investigated the evolutionary properties of the He II glitch in red giants, and demonstrated that under ideal conditions, fitting for it using only the $\ell = 0, 2$ modes \edit2{may not produce substantially} different results from those obtained \edit2{including} the dipole modes. However, the use of dipole modes will naturally improve the robustness of any application of this procedure to observational data, and therefore remains a pressing methodological concern --- especially seeing that present techniques yield uncertainties in the glitch amplitudes too large to be of use as constraints on stellar properties. Future improvements to the technique --- from better treatment of dipole mixed modes, or potentially from further constraints using modes of higher angular degree --- may yet prove to be of diagnostic value. We leave this discussion to potential follow-up work in this direction.

% \begin{acknowledgements}
This research made use of Lightkurve, a Python package for Kepler and TESS data analysis. This work is partially supported by NSF grant AST-2205026 to SB.
\software{\textsc{mesa} \citep{mesa_paper_1, mesa_paper_2, mesa_paper_3, mesa_paper_4, mesa_paper_5}, \textsc{gyre} \citep{townsend_gyre_2013}, \texttt{lightkurve} \citep{2018ascl.soft12013L}, \texttt{astropy} \citep{astropy:2013,astropy:2018}, \texttt{pandas} \citep{pandas}, \texttt{pbjam} \citep{2021AJ....161...62N}, \texttt{yabox} \citep{2017zndo....848679M}}
% \end{acknowledgements}

\def\bibitem{\vskip\bibskip\footnotesize\savebibitem}
\bibliography{biblio.bib}

\end{document}